\documentclass{interact}
\usepackage[english]{babel}
\usepackage[latin1]{inputenc}
\usepackage[T1]{fontenc}
\usepackage{tabularx}
\usepackage{braket}

\usepackage{float}
\usepackage{refstyle}
\usepackage{notes2bib}
\usepackage[sort,compress,super]{natbib}
\usepackage[version=3]{mhchem}
\usepackage{subfig}
\usepackage{amsmath}

\DeclareFontShape{OT1}{cmr}{m}{n}{ <-> cmr12}{}
\usepackage[usenames,dvipsnames]{xcolor}

\newcommand*\citeref[1]{ref.~\citenum{#1}}
\newcommand*\citerefs[1]{refs.~\citenum{#1}}

\newcommand*\osa{\sigma_{\alpha}}
\newcommand*\osb{\sigma_{\beta}}

\newcommand*\si{\sigma}
\newcommand*\de{\delta}
\newcommand*\al{\alpha}
\newcommand*\be{\beta}

\newref{eq}{name = eq.~, names = eqs.~}
\newref{fig}{name = Fig.~, names = Figs.~}
\newref{sec}{name = Section~, names = Sections~}
\newcommand{\etal}{\mbox{\emph{et al.}}}
\newcommand{\eg}{\mbox{\emph{e.g.}}}

\newcommand{\HelFEM}{\textsc{HelFEM}}
\newcommand{\Libxc}{\textsc{Libxc}}
\newcommand{\London}{\textsc{London}}
\newcommand{\Erkale}{\textsc{Erkale}}
\newcommand{\PsiFour}{\textsc{Psi4}}
\newcommand{\Qchem}{\textsc{Q-Chem}}
\newcommand{\half}{\frac{1}{2}}
\newcommand{\vect}[1]{\boldsymbol{#1}}

\begin{document}
\title{Fully numerical electronic structure calculations on diatomic molecules in weak to strong magnetic fields}

\author{ \name{Susi Lehtola,\thanks{CONTACT: Susi Lehtola, email:
      susi.lehtola@alumni.helsinki.fi} Maria Dimitrova,\thanks{Maria
      Dimitrova, email: maria.dimitrova@helsinki.fi} and Dage
    Sundholm\thanks{Dage Sundholm, email: dage.sundholm@helsinki.fi}}
  \affil{Department of Chemistry, University of Helsinki, P.O. Box 55
    (A.  I. Virtasen aukio 1), FI-00014 University of Helsinki,
    Finland} }

\maketitle


\begin{abstract}

We present fully numerical electronic structure calculations on
diatomic molecules exposed to an external magnetic field at the
unrestricted Hartree--Fock limit, using a modified version of a
recently developed finite element program, \HelFEM{}. We have
performed benchmark calculations on a few low-lying states of \ce{H2},
\ce{HeH+}, \ce{LiH}, \ce{BeH+}, \ce{BH}, and \ce{CH+} as a function of
the strength of an external magnetic field parallel to the molecular
axis. The employed magnetic fields are in the range of $B=[0,10]~B_0$
atomic units, where $B_0 \approx 2.35 \times 10^5$ T.  We have
compared the results of the fully numerical calculations to ones
obtained with the \London{} code using a large uncontracted
gauge-including Cartesian Gaussian (GICG) basis set with exponents
adopted from the Dunning aug-cc-pVTZ basis set. By comparison to the
fully numerical results, we find that the basis set truncation error
in the gauge-including Gaussian basis set is of the order of 1
kcal/mol at zero field, that the truncation error grows rapidly when
the strength of the magnetic field increases, and that the largest
basis set truncation error at $B=10~B_0$ exceeds 1000
kcal/mol. Studies in larger Gaussian basis sets suggest that reliable
results can be obtained in GICG basis sets at fields stronger than
$B=B_0$, provided that a sufficient coverage of
higher-angular-momentum functions is included in the basis
set. \end{abstract}


\section{Introduction}
Because of the intrinsically weak strength of the magnetic
interaction, most molecules exposed to even the strongest static
magnetic fields available in laboratory conditions can be studied
computationally with excellent accuracy using perturbation theory. For
example, the most powerful continuous magnetic field obtained under
laboratory conditions is produced by the 45-tesla magnet at the
National High Magnetic Field Laboratory in Florida,\cite{Miller2003}
for which $\mu_B B = 1.1$ meV that is small compared to the thermal
energy $k_B T$, except at very small temperatures $T \lesssim 13$ K.

Stronger fields, however, have to be considered explicitly in
electronic structure calculations, as the magnetic field ceases to be
a small perturbation and becomes of equal or greater importance than
the Coulomb interactions between the electrons. The interest in such
studies arose when Kemp \etal{} discovered circularly polarised light
coming from a white dwarf, and estimated the magnetic field strength
to be about $1000$ T at the surface.\cite{Kemp1970} Zeeman splittings
that are consistent with the lines of light atoms and molecules such
as H, He, O, CH, and \ce{C2} have also been observed in the
atmospheres of magnetic white dwarfs.\cite{Angel1977, Schmidt1995,
  Jordan1998, Wickramasinghe2000, Wickramasinghe2002, Liebert2003} In
addition to these applications in astrophysics, the modeling of
matter in extreme magnetic fields can be beneficial also for other
purposes, as \eg{} the behavior of some solid-state systems with a
high dielectric constant mimics that of atoms in strong magnetic
fields.\cite{Garstang1977, Murdin2013}

However, the computational modeling of the electronic structure of
matter in strong magnetic fields is challenging, in general, since the
magnetic interaction confines the electrons in directions
perpendicular to the field.  Furthermore, since the magnetic field
couples explicitly to the spin, also the ground-state configuration
depends on the field strength, yielding a rich state diagram with
several state crossings. But, in extreme conditions such as in the
atmospheres of highly magnetized neutron stars a.k.a. magnetars where
the field strength can reach billions of tesla,\cite{Duncan1992}
electronic structure calculations become again easier, as the magnetic
interaction becomes dominant and Coulomb interactions can be treated
perturbatively to good accuracy. This is known as the Landau regime in
which the orbitals are elongated along the magnetic field axis due to
the strong perpendicular confinement, and the energy barrier to two-
and three-dimensional molecular structures is so large that only
one-dimensional chains can be formed.\cite{Duncan2000} Several
computational studies on the structure of atoms and chains of atoms in
the strong-field regime have been reported.\cite{Neuhauser1987,
  Lieb1992, Demeur1994, Liberman1995, Relovsky1996, Medin2006,
  Medin2006a, Thirumalai2014}

However, the more interesting and computationally challenging
situations appear when the Coulomb and the magnetic-field interactions
are of similar magnitude; this is called the intermediate
regime,\cite{Lai2001} where the choice of the basis set is
difficult. The traditional linear combination of atomic orbitals
(LCAO) approach, in which the atomic orbitals are by convention
isotropic, works well for zero-field calculations. When the magnetic
interactions become dominant over Coulomb interactions, Landau
orbitals are often employed. However, in the presently investigated
intermediate regime, neither the traditional LCAO expansion nor Landau
orbitals are expected to yield a fast convergence to the basis set
limit.

Gaussian-type orbital (GTO) basis sets are the common choice in
calculations at zero field, because the electron interaction integrals
can be calculated in a GTO basis set in a straightforward
manner. Calculations using GTOs yield good results even at finite
magnetic field strengths, provided that a gauge factor is explicitly
included in the basis set; that is, if gauge-including atomic orbitals
(GIAO) a.k.a. London orbitals are used.\cite{London1937,
  Ditchfield1974, Wolinski1990} Calculations with gauge-including GTOs
have shown that significant changes in molecular structure may be
observed in the intermediate regime:\cite{Schmelcher2012} for
instance, a new type of \emph{paramagnetic bond} can be created by the
interaction of two atoms with high spin multiplicity.\cite{Lange2012}
Lange \etal{} have shown that \ce{H2} exists as a triplet-state
molecule when the magnetic field is stronger than 175~000 T, with the
molecule strongly preferring an orientation perpendicular to the
field.\cite{Lange2012} This paramagnetic bonding mechanism explains
why Xu \etal{} could observe molecular hydrogen in white dwarf
atmospheres,\cite{Xu2013} even though the temperatures therein may
reach over $10~000$ K.

However, the use of conventional isotropic gauge-including GTOs might
lead to large basis set truncation errors at finite fields, as the
electronic structure of atoms deforms continuously from spherical
symmetry at zero field to needle-like structures at strong fields, and
as the latter shape is not readily expandable in such a basis
set. Therefore, finite-field calculations are typically performed only
up to about one atomic unit of field strength $B_0 \approx 2.35 \times
10^5$ T in this kind of basis set. Furthermore, uncontracted basis
sets are used for a better description of the deformation of the
orbitals caused by the magnetic field.\cite{Stopkowicz2015,
  Reynolds2015, Hampe2017, Stopkowicz2018}

Although the problems encountered with isotropic GTO basis sets in the
description of needle-like orbitals can be circumvented by employing
anisotropic GTO basis sets,\cite{Aldrich1979, Schmelcher1988} which
have been successfully used to study atoms\cite{Becken1999,
  Becken2000, Becken2001, Al-Hujaj2004, Al-Hujaj2004a} and
molecules\cite{Detmer1997, Detmer1998, Detmer1998a} in strong magnetic
fields, they have some shortcomings as well. It is more difficult to
optimize their exponents than those of conventional isotropic basis
sets.\cite{Kubo2007, Zhu2014, Zhu2017a} Anisotropic basis sets are
also considerably larger than isotropic ones, because the latter
consist of $N$ GTOs with the same exponents in the three spatial
directions, whereas the former employ a set of $N_\parallel$ and
$N_\perp$ exponents parallel and perpendicular to the field,
respectively, yielding $N_\parallel \times N_\perp$ basis
functions. More general approaches, such as using a full $3\times 3$
matrix of exponents in $(x,y,z)$, are possible as well, which lead to
an even larger number of basis functions. Moreover, the use of
anisotropic exponents requires dedicated software,\cite{Aldrich1979,
  Schmelcher1988, Becken1999} because the calculation of one- and
two-electron integrals becomes more complicated and cannot be
performed with standard molecular integrals codes.

Yet, all the problems inherent in either GTO basis set approaches can
be circumvented by switching to the use of a fully numerical basis
set. Fully numerical approaches typically allow for adaptively refined
basis sets, and a systematical approach to the complete basis set
limit. Even highly anisotropic problems can be readily treated with
such approaches. Furthermore, if a practically complete basis set is
used -- as is typically done in numerical approaches -- the use of
gauge-including basis functions is not necessary for maintaining the
gauge invariance of the magnetic field. For instance, fully numerical
calculations on atoms and \ce{H2+} in weak to strong magnetic fields
have been reported by Ivanov and Schmelcher,\cite{Ivanov1994,
  Ivanov1998, Ivanov1999a, Ivanov2000, Ivanov2001, Ivanov2001a,
  Ivanov2001b} whereas calculations of atoms in magnetar-level
magnetic fields have been reported by Schimeczek
\etal{}\cite{Schimeczek2014} Static magnetic properties employing
linear response at the self-consistent field level of theory have also
been recently reported by Jensen \etal{}\cite{Jensen2016}

Here, we present a fully numerical approach for electronic structure
calculations on diatomic molecules with an explicit, finite magnetic
field along the molecular axis. The matrix elements of the magnetic
field interaction Hamiltonian have been derived and implemented in the
recently published, freely available \HelFEM{} program\cite{HelFEM}
for fully numerical calculations on atoms\cite{Lehtola2019a} and
diatomic molecules.\cite{Lehtola2019b} \HelFEM{} uses a variational
method for solving the self-consistent field equations of
Hartree--Fock (HF) or Kohn--Sham\cite{Kohn1965} density functional
theory (DFT),\cite{Hohenberg1964} and supports hundreds of
exchange-correlation functionals via an interface to the \Libxc{}
library.\cite{Lehtola2018} The present approach is ideal for studies
with magnetic fields strengths belonging to the intermediate regime,
as Coulomb and magnetic interactions are treated self-consistently on
the same footing in \HelFEM{}.

The partial wave expansion originally proposed by
McCullough\cite{McCullough1975, McCullough1986} is used in \HelFEM{}
for calculations on diatomic molecules, in which the ``radial''
direction\cite{Lehtola2019c} is further described with one-dimensional
finite element shape functions; see \citeref{Lehtola2019b} and
\secref{nummeth} for more details. Analogous grid-based approaches
have also been pursued previously; they have been reviewed in
\citeref{Lehtola2019c}. The complete basis set (CBS) limit can be
reached systematically in a variational fashion with the \HelFEM{}
program, as the results converge smoothly and monotonically when the
size of the basis set is increased,\cite{Lehtola2019a, Lehtola2019b}
thus enabling electronic structure studies even at strong magnetic
fields. The prolate spheroidal coordinate system used in the present
work has also been previously used in studies of the \ce{H2+} ion in a
strong magnetic field parallel to the molecular
axis.\cite{Kravchenko1997, Vincke2006}

Numerical calculations on diatomic molecules have a long history of
being used for establishing basis set limits and LCAO basis set
truncation errors in the absence of electromagnetic fields as well as
for finite electric fields, as has been reviewed
elsewhere.\cite{Lehtola2019c} In the present work, we study the
complete basis set (CBS) limits of the singlet, triplet, and quintet
states of \ce{H2}, \ce{HeH+}, \ce{LiH}, \ce{BeH+}, \ce{BH} and
\ce{CH+} in weak to strong magnetic fields by performing fully
numerical calculations with \HelFEM{}. Guided by the CBS limit results
obtained with \HelFEM{}, we assess the accuracy of commonly-used
gauge-including Cartesian Gaussian (CICG) basis sets, as the question
of the reliability of GTO basis sets is an important issue for studies
at finite magnetic fields. The Gaussian-basis calculations are done
with the \London{} program.

The study is undertaken at the unrestricted HF (UHF) level of theory,
which is sufficient for the present purpose of establishing the basis
set truncation error in commonly employed basis sets.  Although both
\HelFEM{} and \London{} also support DFT calculations, the basis set
requirements of HF and DFT are known to be close to identical. UHF
calculations allow for an unbiased comparison of the fully numerical
and GTO approaches.

The organization of the present work is the following.  The basic
theory is briefly presented in \secref{magfield}. The computational
approach is presented in \secref{compmeth}: the numerical approach as
implemented in \HelFEM{} is described in \secref{nummeth}, the
gauge-including isotropic Cartesian Gaussian basis set methods as
implemented in the \London{} code\cite{london} are presented in
\secref{gtometh}, and the details on the calculations are given in
\secref{compdetails}. The results are discussed in \secref{results},
and the main conclusions thereof are summarized in
\secref{summary}. Atomic units are employed, if not specified
otherwise.

\section{Quantum Systems in Finite Magnetic Fields}
\label{sec:magfield}

As is well known from Maxwell's equations, any magnetic field
$\vect{B}$ can be defined in terms of a vector potential $\vect{A}$ as
$ \vect{B} = \nabla \times \vect{A}$. The presence of an
electromagnetic field changes the Hamiltonian for an electron as
\begin{equation}
  H = H_0 + \vect{A}\cdot\vect{p} + \half
  \vect{A}^2, \label{eq:fullHam}
\end{equation}
where $H_0$ is the field-free molecular Hamiltonian and $\vect{p}$ is
the momentum operator which commutes with $\vect{A}$ for a uniform
magnetic field. In the rotational-symmetry preserving symmetric gauge,
the uniform magnetic field $\vect{B}$ is generated by the vector
potential
\begin{equation}
  \vect{A} = \frac 1 2 \vect{B} \times \vect{r}. \label{eq:Auniform}
\end{equation}
\noindent 
However, the vector potential $\vect{A}$ has a gauge degree of freedom
$f(\vect{r},t)$: any potential $\vect{A}' = \vect{A} + \nabla
f(\vect{r},t)$ generates the same magnetic field $\vect{B}$ as
$\vect{A}$. The gauge of $\vect{A}$ can be chosen freely for
calculations, with some choices being more favorable for computation
than others. Typically, the Coulomb gauge
\begin{equation}
  \nabla \cdot \vect{A} = 0 \label{eq:coulombgauge}
\end{equation}
is used, as it leads to the least number of terms in the equations.
This choice still leaves the freedom to choose \eg{} a \emph{gauge
  origin} $\vect{O}$, as replacing $\vect{r}$ with $\vect{r}-\vect{O}$
in \eqref{Auniform} again generates the same field $\vect{B}$. The
inclusion of the gauge origin dependence yields the final expression
for the vector potential
\begin{equation}
    \vect{A}_{\vect{O}}(\vect{r}) = \half \vect{B} \times ( \vect{r} - \vect{O} ) = \half \vect{B} \times \vect{r_O}.
    \label{eq:vecPot}
\end{equation}
Including this dependence on the gauge origin of $\vect{A}$ in the
Hamiltonian, the many-electron Hamiltonian can now be written in terms
of the sums of the orbital and spin angular momenta
\begin{equation}
    \hat{H} = \hat{H}_0 + \frac{1}{2} \sum_i^N \mathbf{B} \cdot
    \mathbf{l}_{\mathbf{O},i} + \mathbf{B} \cdot \mathbf{S} +
    \frac{1}{8} \sum_i^N \left( B^2 r_{\mathbf{O},i}^2 -(\mathbf{B}
    \cdot \mathbf{r}_{\mathbf{O},i})^2\right) \label{eq:Hamvec}
\end{equation}
where $\mathbf{S}$ is the total spin, $ \mathbf{r}_i^O = \mathbf{r}_i
- \mathbf{O} $ is the position of the $i$:th electron with respect to
the global gauge origin $\mathbf{O}$, and $ \mathbf{l}_{O,i} =
-i\mathbf{r}_{O,i} \times \nabla_i $ is the canonical angular
momentum.

Now, choosing $\vect{O} = \vect{0}$ and the magnetic field to coincide
with the $z$ axis as $\vect{B}=(0,0,B)$, \eqref{Hamvec} simplifies to
\begin{equation}
    H = H_0 + \half B L_z + B S_z + \frac{1}{8} B^2 (x^2 + y^2)
    \label{eq:HamZ}
\end{equation}
as then only the $z$ projections of the orbital and spin angular
momentum operators $L_z$ and $S_z$, respectively, remain in the
orbital-Zeeman ($L_z$) and spin-Zeeman ($S_z$) terms. Next, as the
magnetic field direction coincides with the molecular axis, the matrix
elements of \eqref{HamZ} can be made real. The Hamiltonian then has no
explicit dependence on the azimuth angle $\varphi$ around the
molecular axis, and by Noether's theorem the $\varphi$-momentum is
conserved. This leads to a blocking of the molecular orbitals, which
have an azimuth-angular dependence of $\exp(i m \varphi)$ with integer
values of $m$, analogously to the field-free cases discussed in
\citerefs{Lehtola2019a}, \citenum{Lehtola2019b}, and
\citenum{Lehtola2019c}.

The terms in \eqref{Hamvec, HamZ} that are linear in $\vect{B}$
describe paramagnetic response, while the $\vect{B}^2$ terms are
diamagnetic. The paramagnetic terms can either increase or decrease
the energy as the field strength grows, whereas the diamagnetic term
is positive, always increasing the energy of all electronic
systems. All systems become diamagnetic at strong fields due to the
quadratic character of the diamagnetic $\vect{B}^2$
terms.\cite{Tellgren2009} The diamagnetic term leads to a confinement
in the directions perpendicular to the magnetic field, which causes
atoms and molecules to become elongated in the parallel direction in
strong magnetic fields.

Due to the orbital-Zeeman and spin-Zeeman effects, electron
configurations change dramatically with increasing field strength.
High-spin states with occupied $\beta$-spin ($ \downarrow$) orbitals
are favored in strong magnetic fields, and orbitals with large
negative angular momentum quantum numbers are stabilized. For
instance, for atoms in strong magnetic fields, the $f_{-2}$ orbital
can lie energetically much lower than the $s$ orbital of the same
electronic shell.\cite{Stopkowicz2015} Here and in the rest of the
manuscript, we use the sign convention that the energy of orbitals
with $m <0$ and $\beta$ spin are lowered by the external magnetic
field due to the orbital-Zeeman and spin-Zeeman effects, respectively.

\section{Computational Methods}
\label{sec:compmeth}

\subsection{Numerical calculations}
\label{sec:nummeth}

The numerical approach used in \HelFEM{} follows the traditional variational
basis set approach of quantum chemistry; that is, the program solves the
Roothaan--Hall/Pople--Nesbet equations\cite{Roothaan1951, Pople1954}
\begin{equation} \mathbf{F}^\sigma \mathbf{C}^\sigma = \mathbf{S}
\mathbf{C}^\sigma \boldsymbol{\epsilon}^\sigma, \label{eq:roothaan}
\end{equation}
where $\mathbf{F}^\sigma$ and $\mathbf{C}^\sigma$ are the Fock matrix
and molecular orbital coefficients for spin $\sigma$,
$\boldsymbol{\epsilon}^\sigma$ are the corresponding orbital energies,
and $\mathbf{S}$ is the overlap matrix. However, instead of a basis
set composed of atomic orbital basis functions with global support,
\HelFEM{} employs a basis set of finite element shape functions, which
are local; see \citeref{Lehtola2019a} for details on the presently
employed basis functions.  The diatomic approach, in which the
``radial'' functions are expanded in finite-element functions and the
``angular'' part is expanded in spherical harmonics, has been
described in \citeref{Lehtola2019b}. In the present work, \HelFEM{}
has been modified such that the magnetic field interaction terms of
\eqref{HamZ} are added into the core Hamiltonian and Fock matrices, as
is described below. The implementation and calculations otherwise
follow the procedures described in \citerefs{Lehtola2019a} and
\citenum{Lehtola2019b}, where the computational approach is described
in more detail.

Although we consider only molecular calculations in this work, we also
describe the implementation of magnetic fields for atomic calculations
in \HelFEM{}. In both cases, the $z$ component of the angular momentum
is given by
\begin{align}
L_{z}= & i\partial_{\varphi},\label{eq:Lz-def}
\end{align}
which is easily recognized by noting that in either case the $\varphi$
dependence of the orbitals is of the form
$\exp(im\varphi)$,\cite{Lehtola2019a, Lehtola2019b} where $m$ is the magnetic
quantum number.

\subsubsection{Atoms}
\label{sec:numatoms}
The numerical basis set for atomic calculations is\cite{Lehtola2019a}
\begin{equation}
\psi_{nlm}(\boldsymbol{r})=r^{-1}B_{n}(r)Y_{l}^{m}(\theta,\varphi)\label{eq:atorb}
\end{equation}
where $B_n(r)$ are the radial shape functions and
$Y_{l}^{m}(\theta,\varphi)$ are spherical harmonics. The matrix
elements of \eqref{HamZ} are then obtained as
\begin{align}
\braket{i|L_{z}|j}= & \int
B_{i}(r)Y_{l_{i}}^{m_{i}}(\theta,\varphi)^*\left[i\frac{\partial}{\partial\varphi}\right]B_{j}(r)Y_{l_{j}}^{m_{j}}(\theta,\varphi){\rm
  d}\Omega\nonumber \\ = & -m_{j}S_{ij}\label{eq:Lz-ij}
\end{align}
and
\begin{align}
\braket{i|x^{2}+y^{2}|j}= & \int B_{i}(r)r^{2}B_{j}(r){\rm d}r\int\sin^{2} \theta \  Y_{l_{i}}^{m_{i}}(\theta,\varphi)^*Y_{l_{j}}^{m_{j}}(\theta,\varphi){\rm d}\Omega,\label{eq:x2y2-at}
\end{align}
as the parabolic confining term is given by
\begin{equation}
x^{2}+y^{2}=r^{2}\sin^{2}\theta\label{eq:x2y2-atdef}
\end{equation}
in spherical polar coordinates. The only contribution that has to be
added to the expressions in \citeref{Lehtola2019a} is the
$\sin^{2}\theta$ term, which can be handled via Gaunt coefficients
with the expansion
\begin{equation}
\sin^{2}\theta=\frac{4}{3}\sqrt{\pi}Y_{0}^{0}-\frac{4}{15}\sqrt{5\pi}Y_{2}^{0}.\label{eq:sin2}
\end{equation}
For details on the evaluation of the integrals in \eqref{Lz-ij,
  x2y2-atdef}, see \citeref{Lehtola2019a}.

\subsubsection{Diatomic molecules}
\label{sec:numdiat}
For diatomic molecules, the numerical basis set is of the
form\cite{Lehtola2019b}
\begin{equation}
\chi_{nlm}(\mu,\nu,\varphi) = B_{n}(\mu)Y_{l}^{m}(\nu,\varphi),\label{eq:diatorb}
\end{equation}
where $(\mu,\nu,\varphi)$ are transformed prolate spheroidal
coordinates. While $L_z$ maintains the same form as \eqref{Lz-ij},
$\braket{i|L_z|j} = -m_j S_{ij}$, the matrix element for the parabolic
confining term is obtained as
\begin{align}
  \braket{i|x^{2}+y^{2}|j}= & \int B_{i}(\mu)
  Y_{l_{i}}^{m_{i}}(\nu,\varphi)^* \ R_{h}^{2}\sinh^{2}\mu\sin^{2}\nu
  \nonumber \\ \times &
  B_{j}(r)Y_{l_{j}}^{m_{j}}(\nu,\varphi)R_{h}^{3}\sinh\mu\left(\cosh^{2}\mu-\cos^{2}\nu\right){\rm
    d}\Omega\nonumber \\ = &
  R_{h}^{5}\Bigg(\int\sinh^{3}\mu\cosh^{2}\mu
  \ B_{i}(\mu)B_{j}(\mu){\rm d}\mu\int\sin^{2}\nu
  \ Y_{l_{i}}^{m_{i}}(\nu,\varphi)^*Y_{l_{j}}^{m_{j}}(\nu,\varphi)d\Omega\nonumber
  \\ & -\int\sinh^{3}\mu \ B_{i}(\mu)B_{j}(\mu){\rm
    d}\mu\int\sin^{2}\nu\cos^{2}\nu
  \ Y_{l_{i}}^{m_{i}}(\nu,\varphi)^*Y_{l_{j}}^{m_{j}}(\nu,\varphi)d\Omega\Bigg)\label{eq:x2y2-diat}
\end{align}
which can again be evaluated using \eqref{sin2} and the additional
expansion
\begin{equation}
\sin^{2}\nu\cos^{2}\nu=\frac{4}{15}\sqrt{\pi}Y_{0}^{0}+\frac{4}{105}\sqrt{5\pi}Y_{2}^{0}-\frac{16}{105}\sqrt{\pi}Y_{4}^{0}.\label{eq:sin2cos2}
\end{equation}
For details on the evaluation of the integrals in $\langle i | L_z |
j\rangle$ and \eqref{x2y2-diat}, see \citeref{Lehtola2019b}.

\subsection{Gaussian basis set calculations}
\label{sec:gtometh}

Gauge-independent electronic structure LCAO calculations on atoms and
molecules exposed to an external magnetic field can be performed by
modifying the basis set to include an explicit gauge transformation
factor\cite{London1937, Ditchfield1974, Wolinski1990}
\begin{equation}
    \psi_{nlm}^\text{GIAO} (\mathbf{r}) = \exp \left[ \frac i 2
      \vect{B} \times (\vect{G} - \vect{O}) \cdot \vect{r} \right]
    \psi_{nlm} (\mathbf{r}),
    \label{eq:psiGauge}
\end{equation}
resulting in GIAOs, or London orbitals. The complex prefactor in
\eqref{psiGauge} corresponds to a gauge-origin transformation from
$\vect{O}$ to $\vect{G}$ (\eqref{vecPot}), which makes the wave
function invariant to first-order changes in the gauge origin.

The \London{} program employs GIAOs formed of Cartesian
GTOs.\cite{Tellgren2008} A primitive Cartesian GTO centered at
$(x_a,y_a,z_a)$ is given by
\begin{equation}
    \psi_{a} (\mathbf{r}) = N_a (x-x_a)^{i_a} (y-y_a)^{j_a}
    (z-z_a)^{k_a} \exp \left( -\alpha_a r_a^2 \right)
    \label{eq:GTO}
\end{equation}
where $i_a,j_a,k_a$ are integers, the sum of which corresponds to the
angular character of the shell (\eg{} $i_a+j_a+k_a=0,1,2,3$ for $s$,
$p$, $d$, and $f$ shells, respectively), $\alpha_a$ is the exponent,
$N_a$ is the corresponding normalization factor, and $r_a^2 =
(x-x_a)^2 + (y-y_a)^2 + (z-z_a)^2$ is the square of the distance from
the center of the basis function, which typically resides at a
nucleus.

\subsection{Computational details}
\label{sec:compdetails}

We have studied the total energy of a few low-lying states of six
diatomic molecules as a function of the strength of the external
magnetic field. Since the aim of this study is to assess the quality
of calculations employing commonly-used Gaussian basis sets, the
unrestricted Hartree--Fock level of theory is sufficient, and fixed
internuclear distances ($R$) have been employed. The studied molecules
include \ce{H2} ($R=1.4~ a_0$), \ce{HeH+} $(R=1.5~ a_0)$, \ce{LiH}
($R=3.0~ a_0$), \ce{BeH+} ($R=2.5~ a_0$), \ce{BH} $(R=2.3~ a_0$), and
\ce{CH+} ($R=2.1~ a_0$) in their singlet, triplet, and quintet spin
states; the internuclear distances ($R$) were chosen based on the
equilibrium bond length of the molecular species for $B=0$ at the
HF/def2-SVP level of theory.\cite{Weigend2005} Correlation effects or
the fact that the equilibrium bond distances shrink with increasing
strength of the magnetic field along the molecular axis have not been
considered in this work.

As was already stated in \secref{gtometh}, the GTO calculations were
performed with the \London{} program developed by Tellgren
\etal{}\cite{london, Tellgren2008} An uncontracted gauge-including
Cartesian basis set was used, with isotropic exponents adopted from
the triple-$\zeta$ correlation consistent basis set augmented with
diffuse functions (aug-cc-pVTZ).\cite{Dunning1989, Woon1995} This
basis set was chosen for the present work, as it has also been used in
previous studies.\cite{Lange2012} A linear dependency threshold of
$10^{-8}$ was used in the canonical basis set orthonormalization
procedure. The direct inversion of the iterative subspace (DIIS)
method\cite{Pulay1980, Pulay1982} was used for accelerating the
convergence of the self-consistent field procedure. An orbital
rotation convergence threshold of $10^{-6} E_h$ was used in the
\London{} calculations. In problematic cases, the lowest state was
obtained by restarting the calculations from converged density
matrices for the desired configuration for a different field strength.

The fully numerical calculations were carried out using the \HelFEM{}
program, which employs finite-element basis functions, as was detailed
above in \secref{nummeth}. The DIIS method was likewise used in the
\HelFEM{} calculations, in combination with the ADIIS
algorithm.\cite{Hu2010} An orbital rotation convergence threshold of
$10^{-7} E_h$ was used in the \HelFEM{} calculations.

For simplicity, the same numerical basis set was used for all
molecules. The numerical basis set was determined at zero field by the
proxy approach described in \citeref{Lehtola2019b}. Taking the largest
basis set parameters for each molecule, it was found that energies
beyond microhartree accuracy at zero field are obtained for all
molecules with a linear element grid containing three 15-node radial
Lobatto elements, with $l_\sigma=15$ and $l_\pi=11$.  A closer study
showed that \ce{BH} and \ce{CH+} have a $\Delta$ ground state for some
field strengths. The initial basis set was therefore increased to
include $\delta$ orbitals with $l_\delta=7$, which was determined by
extrapolating from $l_\sigma$ and $l_\pi$.

Although magnetic flux density is traditionally measured in units of
tesla, in computational studies the field strength is instead
expressed in atomic units $\beta = B/B_0$. One atomic unit of magnetic
flux density $B_0=E_\mathrm{h} / ea_0 c\alpha$ is equal to $B_0 =
2.350~517~42(20) \times 10^{5}$ T, where $E_\mathrm{h}=
4.359~744~17(75) \times 10^{-18}$ J, $e = 1.602~176~53(14)\times
10^{-19}$ C, $c =299~792~458$ m/s and the inverse fine structure
constant is $\alpha^{-1}=4\pi\epsilon_0 \hbar c / e^2
=137.035~999~11(46)$.

The investigated magnetic field strengths are in the range of
$B=0-10~B_0$ in steps of $0.1~ B_0$. The GTO results obtained for
magnetic fields larger than $B \gtrsim B_0$ may not be entirely
reliable, because the employed basis set is not able to span the
orbitals accurately at strong fields. The extreme case $B = 10~ B_0$
is certainly beyond the scope of the isotropic GTO approach studied in
the present work. Due to the prevalence of state crossings in the
calculations with magnetic field, we have explicitly tracked several
electronic configurations for the studied molecules by fixing the
symmetry of the occupied orbitals in the fully numerical calculations,
as described in \citerefs{Lehtola2019a} and
\citenum{Lehtola2019b}. The ground state energy has then been
extracted by identifying the lowest energy at each value of the
magnetic field strength.

In order to study the basis set convergence at finite magnetic field,
larger numerical basis sets were formed by increasing the number of
partial waves by $4$ simultaneously in each $m$ subchannel, whereas
five radial elements were used in all calculations. Although the
accuracy of the fully numerical calculations was found to decrease
monotonically with increasing field strength, the basis set
convergence was found to be exponential with the basis set size. While
the numerical calculations are fully converged to the basis set limit
at $B=0$, calculations with the largest basis set consisting of five
radial elements and $l_\sigma=31$, $l_\pi=27$, $l_\delta=23$ are
estimated to have millihartree accuracy at $B=10~ B_0$, which is
sufficient for the purpose of the present study.

\section{Results}
\label{sec:results}

\subsection{General results}

\begin{figure}

  \subfloat[\ce{H2}]{\includegraphics[width=0.33\textwidth]{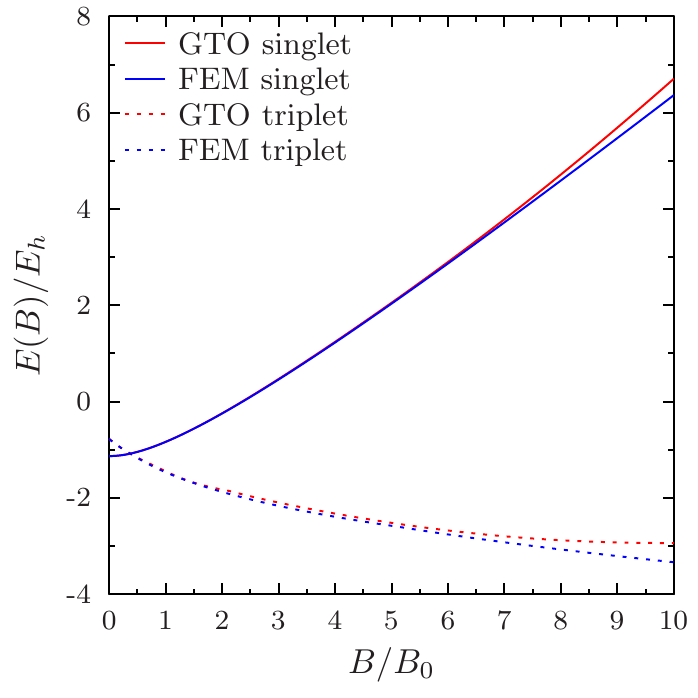}}
  \subfloat[\ce{HeH+}]{\includegraphics[width=0.33\textwidth]{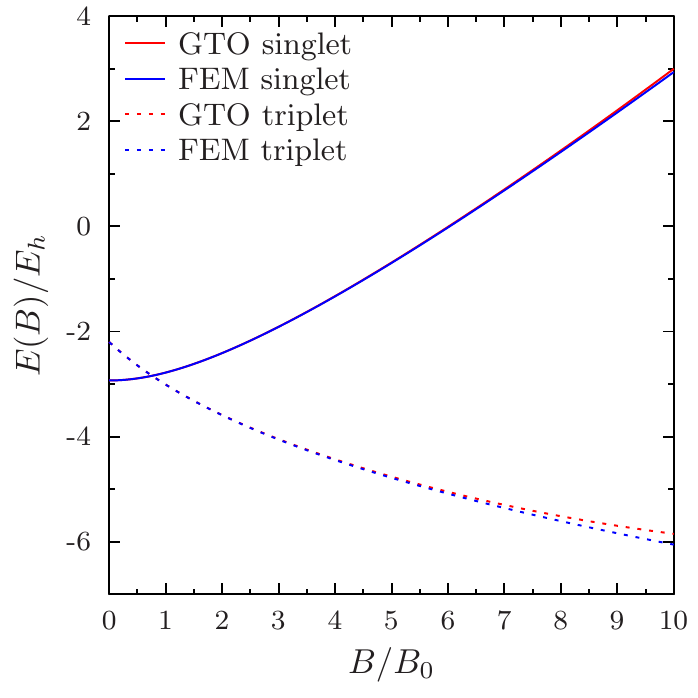}}
  \subfloat[\ce{LiH}]{\includegraphics[width=0.33\textwidth]{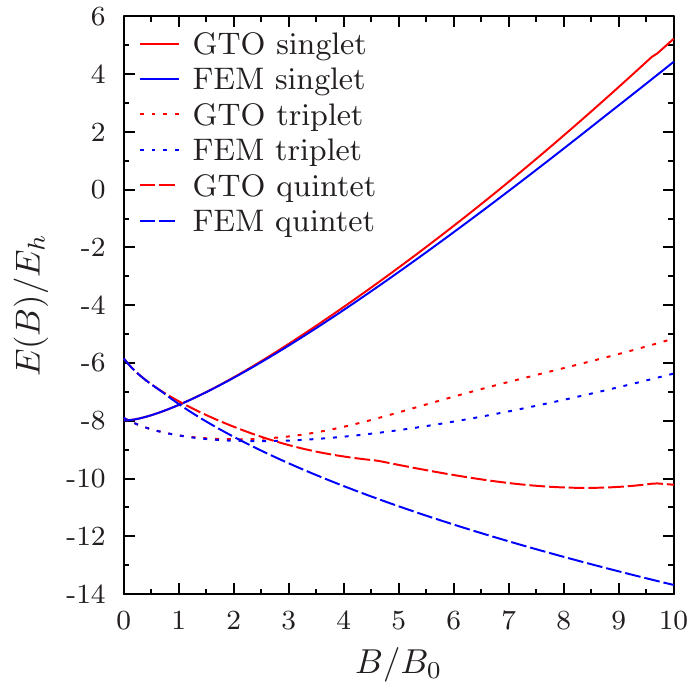}}

  \subfloat[\ce{BeH+}]{\includegraphics[width=0.33\textwidth]{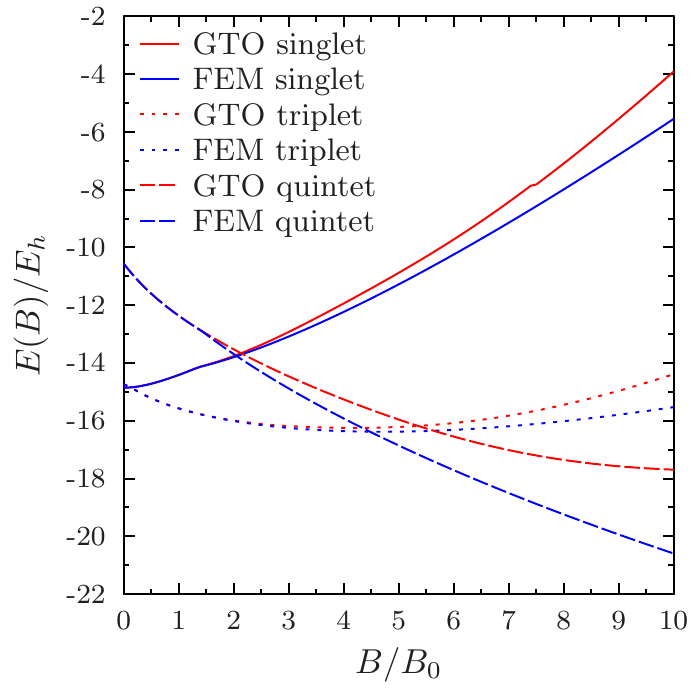}}
  \subfloat[\ce{BH}]{\includegraphics[width=0.33\textwidth]{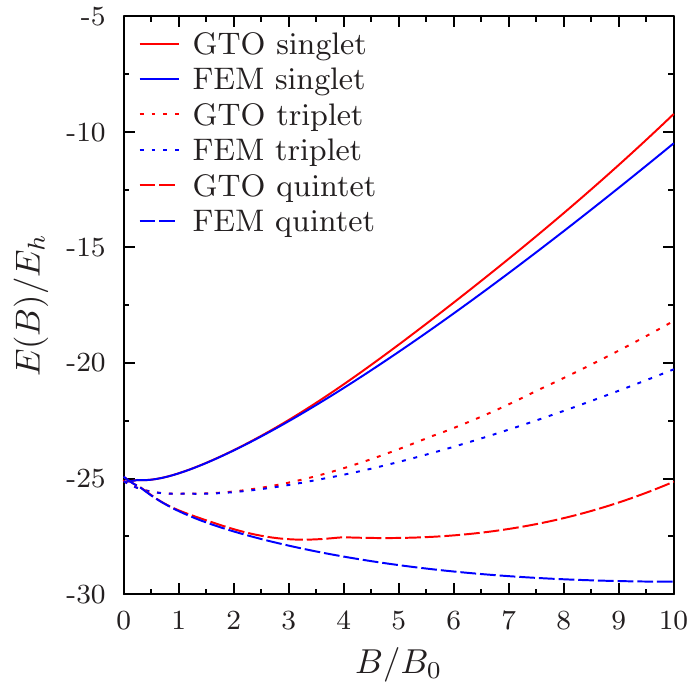}}
  \subfloat[\ce{CH+}]{\includegraphics[width=0.33\textwidth]{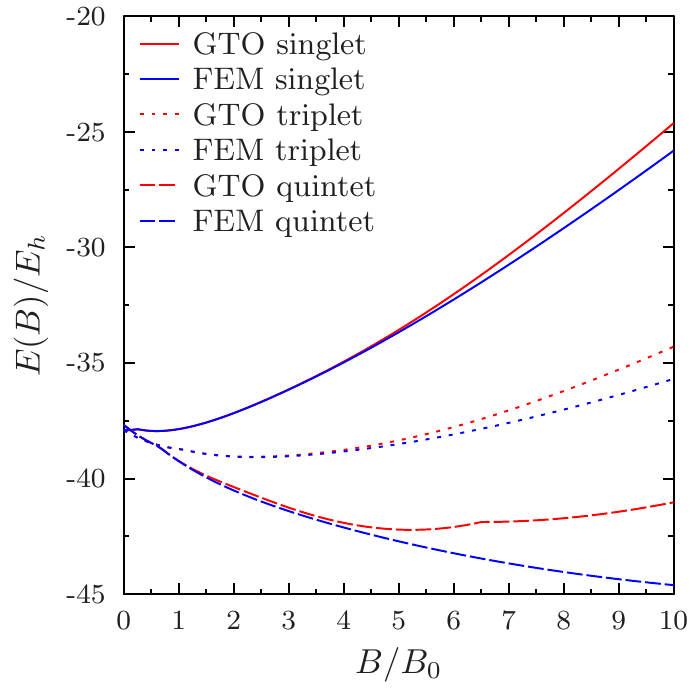}}

\caption{The total energy of \ce{H2}, \ce{HeH+}, \ce{LiH}, \ce{BeH+},
  \ce{BH}, and \ce{CH+} as a function of the strength of the magnetic
  field calculated at the Hartree--Fock level using gauge-including
  Gaussian-type orbital (GTO) and finite element (FEM) basis sets,
  respectively.\label{fig:absE}}

\end{figure}

\begin{figure}
  \subfloat[\ce{H2}]{\includegraphics[width=0.33\textwidth]{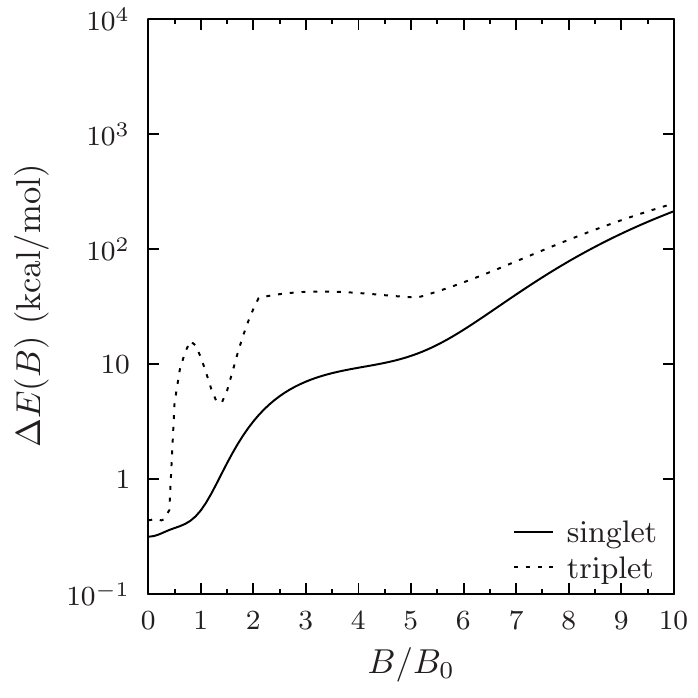}}
  \subfloat[\ce{HeH+}]{\includegraphics[width=0.33\textwidth]{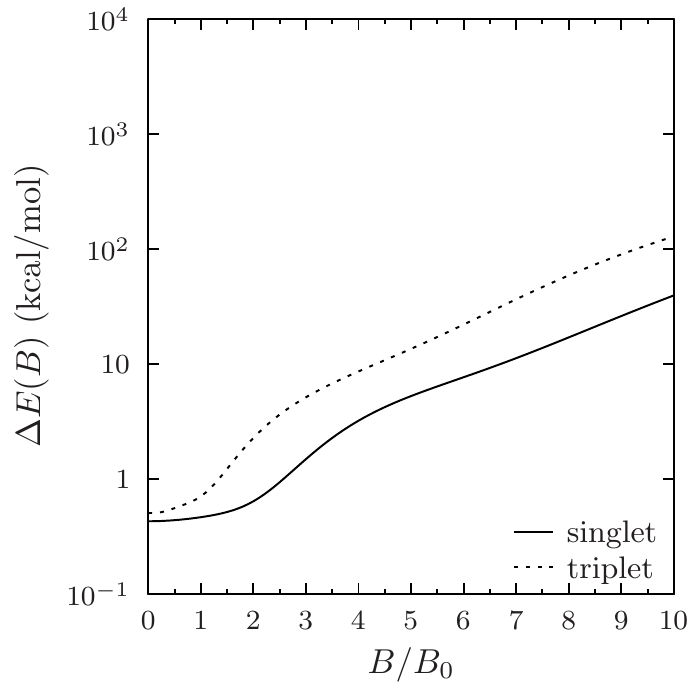}}
  \subfloat[\ce{LiH}]{\includegraphics[width=0.33\textwidth]{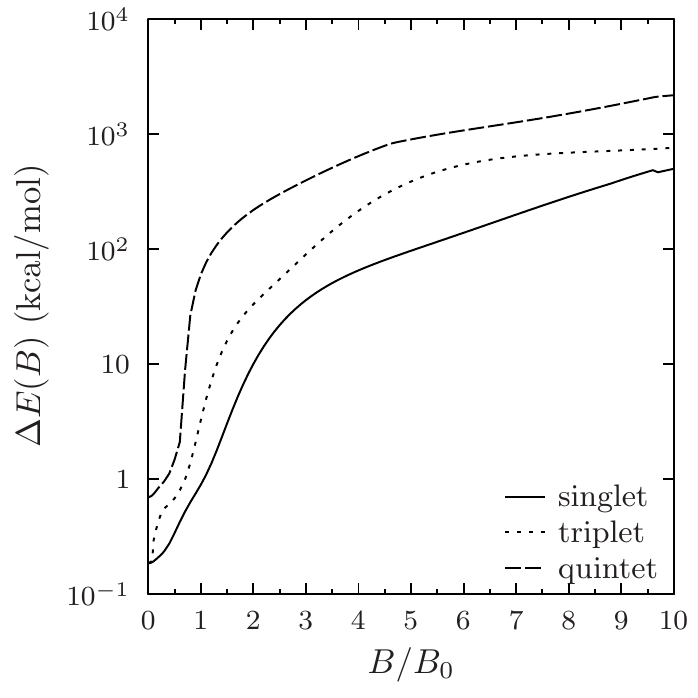}}

  \subfloat[\ce{BeH+}]{\includegraphics[width=0.33\textwidth]{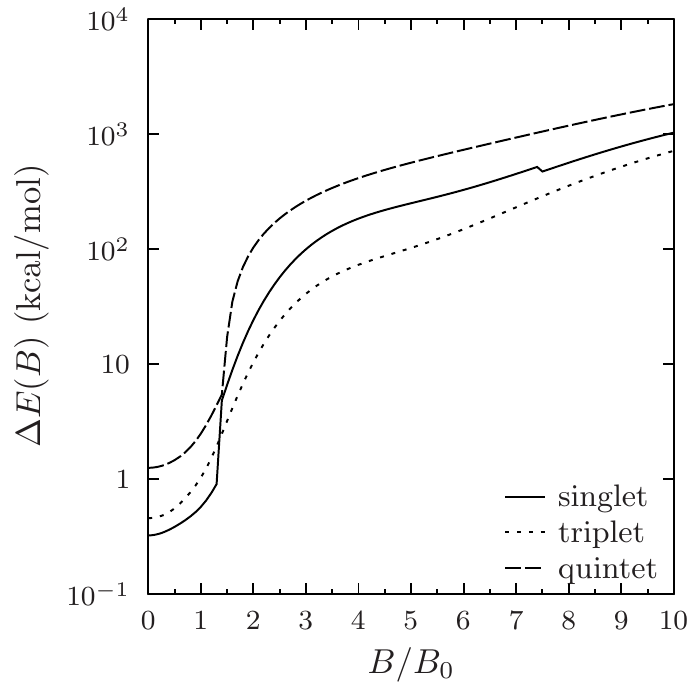}}
  \subfloat[\ce{BH}]{\includegraphics[width=0.33\textwidth]{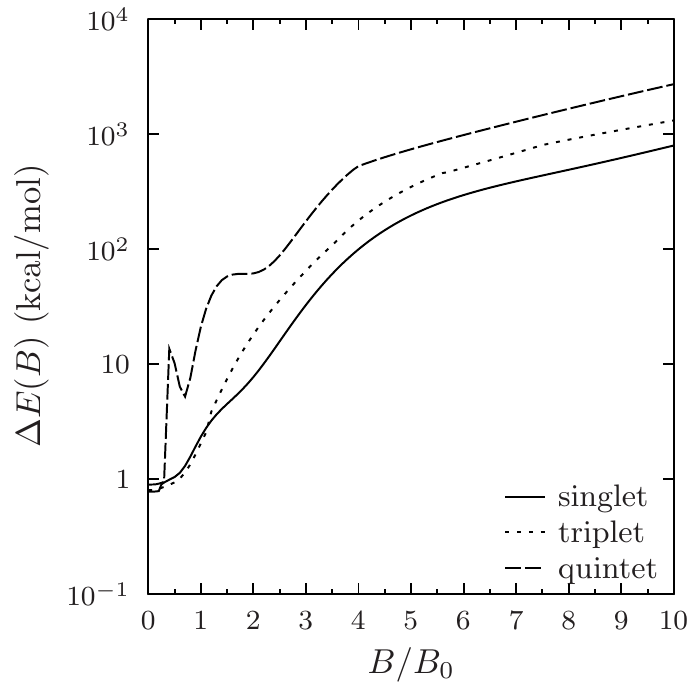}}
  \subfloat[\ce{CH+}]{\includegraphics[width=0.33\textwidth]{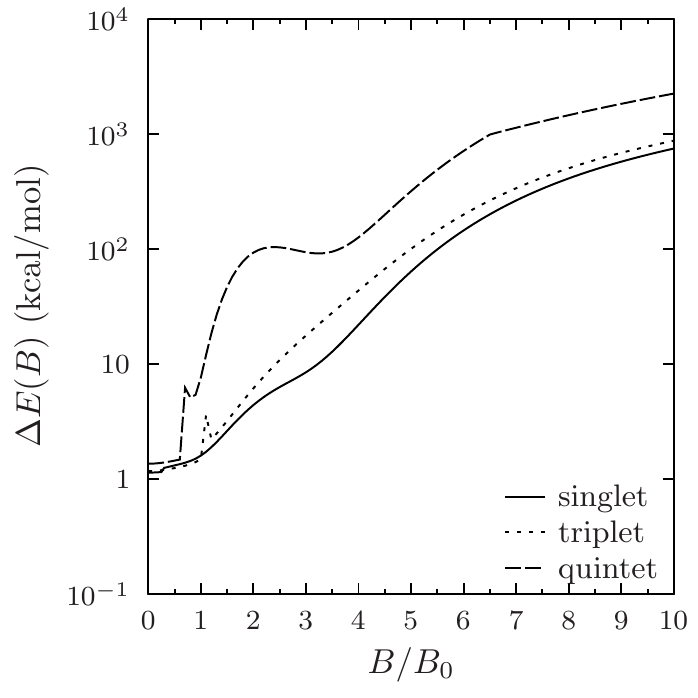}}

\caption{The basis set truncation errors as a function of the magnetic
  field strength, which are estimated as the difference in the
  Hartree--Fock energies obtained in calculations with gauge-including
  Gaussian-type orbitals (GTO) and finite element (FEM) basis sets,
  respectively. Note that the error in energy on the $y$ axis has a
  logarithmic scale.\label{fig:diffE} }
\end{figure}

Total energies as a function of the magnetic field strength calculated
using GTO and numerical basis sets are shown in \figref{absE}. The
obtained ground state configurations are given in \tabref{gs-B-a} for
\ce{H2}, \ce{HeH+}, LiH, in \tabref{gs-B-b} for \ce{BeH+}, and in
\tabref{gs-B-c} for BH and \ce{CH+}. The GTO basis set truncation
errors in \figref{diffE} were obtained by comparing the total energies
of the GTO calculations with \London{} with the converged fully
numerical values from with the \HelFEM{} code.

Although many states have been explicitly considered in the present
work, the data in \figref{absE, diffE} correspond only to the lowest
energy at each value of the field. Plots of the energies of the
individual configurations are shown in the Supporting Information, in
addition to detailed data on how the energy ordering of the $\sigma$,
$\pi$, and $\delta$ orbitals changes with magnetic field.

The curves in \figref{diffE} display oscillations, which can
tentatively be attributed to the GTO basis set.  As the same basis set
is used for all field strengths, the basis set truncation error can be
smaller for some $B$ values, for which the GTO expansion is better
able to describe the electronic wave function than for other values of
$B$. This basis set artifact can result in the observed oscillatory
behavior.

The basis set errors of the GTO calculations at zero field are of the
order of 1 kcal/mol (1 eV = 23.06054801(14) kcal/mol) with the
employed gauge-including fully uncontracted Cartesian triple-$\zeta$
basis set, but the accuracy decays rapidly with increasing field
strength. The poor performance of GTO basis sets for absolute energies
is well-known, whereas relative energies are usually more accurate due
to cancellation of errors.  However, in the presence of strong
magnetic fields, relative energies such as spin-state splittings may
also become inaccurate, since the accuracy of the GTO calculations
varies strongly with the spin state, as will be seen below.

\begin{table}
\caption{Ground-state configurations for \ce{H2}, \ce{HeH+}, LiH, and
  \ce{BeH+} as a function of the magnetic field strength calculated at
  the Hartree--Fock level using gauge-including Cartesian forms of the
  uncontracted aug-cc-pVTZ basis set of Gaussian-type orbitals (GTO),
  and with the finite element (FEM) basis set.  The label X refers to
  symmetry-broken orbitals.  Note that the orbitals are not sorted
  according to their energies.
    \label{tab:gs-B-a}}
\begin{tabularx}{\linewidth}{llllc}
\hline
\hline
Molecule & spin state & basis & configuration & field strength ($B_0$) \\
\hline
    \ce{H2} & singlet& FEM & $ 1\osa 1\osb $ & $ 0 \leq B \leq 10 $
    \vspace{3pt} \\
     & & GTO & $ 1\osa 1\osb $ & $ 0 \leq B \leq 10 $
     \vspace{5pt} \\
    & triplet & FEM & $ (1\si 2\si)_\be $ & $ 0 \leq B < 0.5 $ \\
    & & & $ (1\si 1\pi)_\be $ & $ 0.5 \leq B < 5.2 $ \\
    & & & $ (1\si 2\si)_\be  $ & $ 5.2 \leq B \leq 10 $
    \vspace{3pt} \\
    & & GTO & $ (1\si 2\si)_\be $ & $ 0 \leq B < 0.9 $ \\
    & & & $ (1\si 1\pi)_\be $ & $ 0.9 \leq B < 2.2 $ \\
    & & & $ (1\si 2\si)_\be $ & $ 2.2 \leq B \leq 10 $ \\
\\
    \ce{HeH+} & singlet & FEM & $ 1\osa 1\osb $ & $ 0 \leq B \leq 10 $
    \vspace{3pt} \\
    & & GTO & $ 1\osa 1\osb $ & $ 0 \leq B \leq 10 $
    \vspace{5pt} \\
    & triplet & FEM & $ (1\si 2\si)_\be $ & $ 0 \leq B \leq 10 $
    \vspace{3pt} \\
    & & GTO & $ (1\si 2\si)_\be $ & $ 0 \leq B \leq 10 $
    \vspace{3pt} \\
\\
    LiH & singlet & FEM & $ (1\si 2\si)_\al (1\si 2\si)_\be $ & $0 \leq B < 8.8 $ \\
    & & & $ (1\si 1\pi)_\al (1\si 1\pi)_\be $ & $ 8.8 \leq B \leq 10 $
    \vspace{3pt} \\
    & & GTO & $ (1\si 2\si)_\al (1\si 2\si)_\be $ & $0 \leq B < 9.6 $ \\
    & & & $ (1\si X)_\al (1\si X)_\be $ & $ 9.6 \leq B \leq 10 $
    \vspace{5pt} \\
    & triplet & FEM &  $ 1\osa (1\si 2\si 3\si)_\be $ & $ 0 \leq B < 0.1 $ \\
    & & & $ 1\osa (1\si 2\si 1\pi)_\be $ & $ 0.1 \leq B \leq 10 $
    \vspace{3pt} \\
    & & GTO &  $ 1\osa (1\si 2\si 3\si)_\be $ & $ 0 \leq B < 0.1 $ \\
    & & & $ 1\osa (1\si 2\si 1\pi)_\be $ & $ 0.1 \leq B \leq 10 $
    \vspace{5pt} \\
    & quintet & FEM & $ (1\si 2\si 3\si 1\pi)_\be $ & $ 0  \leq B < 0.7  $ \\
    & & & $ (1\si 2\si 1\pi 1\de)_\beta $ & $ 0.7  \leq B \leq 10  $ 
    \vspace{3pt} \\
    & & GTO & $ (1\si 2\si 3\si 4\si)_\be $ & $ 0 \leq B < 0.01 $ \\
    & & & $ (1\si 2\si 3\si 1\pi)_\be $ & $ 0.01 \leq B < 4.7 $ \\
    & & & $ (1\si 2\si 1\pi 2\pi)_\be $ & $ 4.7 \leq B < 9.7 $ \\
    & & & $ (1\si 2\si 3\si 1\pi)_\be $ & $ 9.7 \leq B \leq 10 $ \\
\hline
\hline
\end{tabularx}
\end{table}

\begin{table}
  \caption{Ground-state configurations for \ce{BeH+} as a function of
    the magnetic field strength calculated at the Hartree--Fock level
    using gauge-including Cartesian forms of the uncontracted
    aug-cc-pVTZ basis set of Gaussian-type orbitals (GTO), and with
    the finite element (FEM) basis set. Note that the orbitals are not
    sorted according to their energies.
    \label{tab:gs-B-b}}
  \begin{tabularx}{\linewidth}{llllc}
    \hline
    \hline
    Molecule & spin state & basis & configuration & field strength ($B_0$) \\
    \hline
    \ce{BeH+} & singlet & FEM & $ (1\si 2\si)_\al (1\si 2\si)_\be $ & $ 0 \leq B < 1.4 $ \\
    & & & $ (1\si 1\pi)_\al (1\si 1\pi)_\be $ & $ 1.4 \leq B \leq 10 $
    \vspace{3pt} \\
    & & GTO & $ (1\si 2\si)_\al (1\si 2\si)_\be $ & $ 0 \leq B < 1.4 $ \\
    & & & $ (1\si 1\pi)_\al (1\si 1\pi)_\be $ & $ 1.4 \leq B \leq 10 $ \\
    \vspace{5pt} \\
    & triplet & FEM &  $ 1\osa (1\si 2\si 3\si)_\be $ & $ 0 \leq B < 0.2 $ \\
    & & & $ 1\osa (1\si 2\si 1\pi)_\be $ & $ 0.2 \leq B \leq 10 $
    \vspace{3pt} \\
    & & GTO &  $ 1\osa (1\si 2\si 3\si)_\be $ & $ 0 \leq B \leq 0.2 $ \\
    & & &  $ 1\osa (1\si 2\si 1\pi)_\be $ & $ 0.2 \leq B \leq 10 $ \\
    \vspace{5pt} \\
    & quintet & FEM & $ (1\si 2\si 3\si 1\pi)_\be $ & $ 0  \leq B < 1.5  $ \\
    & & & $ (1\si 2\si 1\pi 1\de)_\be $ & $ 1.5  \leq B \leq 10  $ \\
    \vspace{3pt} \\
    & & GTO & $ (1\si 2\si 3\si 1\pi)_\be $ & $ 0 \leq B \leq 10 $ \\
\hline
\hline
\end{tabularx}
\end{table}

\begin{table}
\caption{Ground-state configurations for \ce{BH} and \ce{CH+} as a
  function of the magnetic field strength calculated at the
  Hartree--Fock level using gauge-including Cartesian forms of the
  uncontracted aug-cc-pVTZ basis set of Gaussian-type orbitals (GTO),
  and with the finite element (FEM) basis set. ${\pi}_{+}$ orbitals have a higher energy
  in the presence of the magnetic field than at zero field.  Note that
  the orbitals are not sorted according to their
  energies.  \label{tab:gs-B-c}}

\begin{tabularx}{\linewidth}{llllc}
\hline
\hline
Molecule & spin state & basis & configuration & field strength ($B_0$) \\
\hline
    \ce{BH} & singlet & FEM & $ (1\si 2\si 3\si)_\al (1\si 2\si 3\si)_\be $ & $ 0 \leq B < 0.3 $ \\
    & & & $ (1\si 2\si 1\pi)_\al (1\si 2\si 1\pi)_\be $ & $ 0.3 \leq B \leq 10 $
    \vspace{3pt} \\
    & & GTO & $ (1\si 2\si 3\si)_\al (1\si 2\si 3\si)_\be $ & $ 0 \leq B < 0.3 $ \\
    & & & $ (1\si 2\si 1\pi)_\al (1\si 2\si 1\pi)_\be $ & $ 0.3 \leq B \leq 10 $ \\
    \vspace{5pt} \\
    & triplet & FEM & $ (1\si 2\si)_\al (1\si 2\si 3\si 1\pi)_\be $ & $ 0 \leq B < 1.2 $ \\
    & & & $ (1\si 1\pi)_\al (1\si 2\si 3\si 1\pi)_\be $ & $ 1.2 \leq B < 3.6 $ \\
    & & & $ (1\si 1\pi)_\al (1\si 2\si 1\pi 2\pi)_\be $ & $ 3.6 \leq B \leq 10 $
    \vspace{3pt} \\
    & & GTO & $ (1\si 2\si)_\al (1\si 2\si 3\si 1\pi)_\be $ & $ 0 \leq B < 1.5 $ \\
    & & & $ (1\si 1\pi)_\al (1\si 2\si 3\si 1\pi)_\be  $ & $ 1.5 \leq B \leq 10 $ \\
    \vspace{5pt} \\
    & quintet & FEM & $ 1\osa (1\si 2\si 3\si 1\pi 1{\pi}_{+})_\be $ & $ 0 \leq B < 0.4 $ \\
    & & & $ 1\osa (1\si 2\si 3\si 1\pi 1\de)_\be $ & $ 0.4 \leq B \leq 10 $  \\
    \vspace{3pt} \\
    & & GTO & $ 1\osa (1\si 2\si 3\si 1\pi 1{\pi}_{+})_\be $ & $ 0 \leq B < 0.4 $ \\
    & & & $ 1\osa (1\si 2\si 3\si 4\si 1\pi)_\be $ & $ 0.4 \leq B < 0.7 $  \\
    & & & $ 1\osa (1\si 2\si 3\si 1\pi 2\pi)_\be $ & $ 0.7 \leq B < 1.5 $  \\
    & & & $ 1\osa (1\si 2\si 3\si 1\pi 1\de)_\be $ & $ 1.5 \leq B < 4.0 $  \\
    & & & $ 1\osa (1\si 2\si 3\si 4\si 1\pi)_\be $ & $ 4.0 \leq B \leq 10 $  \\
\\
    \ce{CH+} & singlet & FEM & $ (1\si 2\si 3\si)_\al (1\si 2\si 3\si)_\be $ & $ 0 \leq B < 0.3 $ \\
    & & & $ (1\si 2\si 1\pi)_\al (1\si 2\si 1\pi)_\be  $ & $ 0.3 \leq B \leq 10 $
    \vspace{3pt} \\
    & & GTO & $ (1\si 2\si 3\si)_\al (1\si 2\si 3\si)_\be  $ & $ 0 \leq B < 0.3 $ \\
    & & & $ (1\si 2\si 1\pi)_\al (1\si 2\si 1\pi)_\be $ & $ 0.3 \leq B \leq 10 $ \\
    \vspace{5pt} \\
    & triplet & FEM & $ (1\si 2\si)_\al (1\si 2\si 3\si 1\pi)_\be $ & $ 0 \leq B < 1.1 $ \\
    & & & $ (1\si 1\pi)_\al (1\si 2\si 3\si 1\pi)_\be $ & $ 1.1 \leq B \leq 10 $
    \vspace{3pt} \\
    & & GTO & $ (1\si 2\si)_\al (1\si 2\si 3\si 1\pi)_\be $ & $ 0 \leq B < 1.3 $ \\
    & & & $ (1\si 1\pi)_\al (1\si 2\si 3\si 1\pi)_\be $ & $ 1.3 \leq B \leq 10 $ \\
    \vspace{5pt} \\
    & quintet & FEM & $ 1\osa (1\si 2\si 3\si 1\pi 1{\pi}_{+})_\be  $ & $ 0 \leq B < 0.7 $ \\
    & & & $ 1\osa (1\si 2\si 3\si 1\pi 1\de)_\be $ & $ 0.7 \leq B \leq 10 $
    \vspace{3pt} \\
    & & GTO & $ 1\osa (1\si 2\si 3\si 1\pi 1{\pi}_{+})_\be $ & $ 0 \leq B < 1.2 $ \\
    & & & $ 1\osa (1\si 2\si 3\si 1\pi 1\de)_\be $ & $ 1.2 \leq B < 6.5 $ \\
    & & & $ 1\osa (1\si 2\si 3\si 4\si 1\pi)_\be $ & $ 6.5 \leq B \leq 10 $ \\
\hline
\hline
\end{tabularx}
\end{table}

\subsection{Singlet states}

The difference of the GTO and the numerical energies is 0.5--2.3
kcal/mol at $ B = B_0 $ for all studied singlet states.  The GTO
truncation errors approach 1000 kcal/mol at strong magnetic fields for
all but the lightest molecules, \ce{H2} and \ce{HeH+}.

The ground-state orbital occupations of the \London{} and \HelFEM{}
calculations agree well for the singlet states. The problematic cases
turned out to be \ce{LiH} and \ce{BeH+}. \ce{H2} and \ce{HeH+} have
the $ 1\si_\al 1\si_\be$ electron configuration in the investigated
magnetic field strength of $[0,10]~B_0$.  In the GTO calculations, the
energy of the singlet state of \ce{LiH} has a kink at $B = 9.6~ B_0$,
because the electron configuration changes from $(1\si 1\si)_\al (1\si
1\si)_\be$ to a state where the highest occupied $\alpha$ and $\beta$
spin-orbitals have neither $\sigma$ nor $\pi$ symmetry. In contrast,
the \HelFEM{} calculations employ axial symmetry that prevents
symmetry breaking. In the fully numerical calculations on \ce{LiH},
the orbital configuration changes from $(1\si 2\si)_\al (1\si
2\si)_\be$ to $(1\si 1\pi)_\al (1\si 1\pi)_\be$ at $B=8.8~ B_0$. This
configuration was not obtained in the \London{} calculations.

Symmetry breaking is a well-known artifact of the Hartree--Fock method
that originates from the neglect of electron correlation. A full wave
function description includes all configurations which may be
spatially dissimilar, the superposition of which, however, has the
same symmetry as the system itself. As Hartree--Fock considers only
one of these configurations, which does not reflect the full symmetry
of the system, it is sometimes advantageous to further break the
symmetries that are fulfilled by the exact wave function. Because the
character of the symmetry-broken GTO calculation does not match that
of the numerical calculations, the data corresponding to
symmetry-broken configurations have been excluded from the comparison.

Calculations on the singlet state of \ce{BeH+} yield the same orbital
occupation with both methods.  The crossing between the $(1\si
2\si)_\al (1\si 2\si)_\be$ and the $(1\si 1\pi)_\al (1\si 1\pi)_\be$
states occurs at $B = 1.4~ B_0$ -- significantly earlier than for
\ce{LiH}.  The GTO calculations yielded a symmetry-broken state at $B
= 7.7~ B_0$. By tracing the symmetry-broken solution down to weaker
fields, it turned out to be lower in energy than the state with the
$(1\si 2\si)_\alpha (1\si 1 \pi)_\beta$ configuration all the way down
to $B = 0.8~ B_0$, where it crosses the $(1\si 2\si)_\alpha (1\si
2\si)_\beta$ state. As the symmetry-broken state is again a
Hartree--Fock artifact, we did not consider it further and excluded it
from the graphs.

The singlet state of \ce{BH} has a $(1\si 2\si 3\si)_\al (1\si 2\si
3\si)_\be$ configuration at weak fields. Both the \London{} and
\HelFEM{} calculations predict that a state crossing occurs at $B =
0.3~ B_0$, where a $\pi$ orbital becomes the highest occupied orbital.

The \ce{CH+} cation has a weak-field ground state with occupied
orbitals of only $\sigma$ symmetry.  For field strengths $B \geq 0.3~
B_0$, the ground-state configuration becomes $ (1\si 2\si 1\pi)_\al
(1\si 2\si 1\pi)_\be$, as for \ce{BH}. Symmetry analysis of the
\London{} calculations showed that the $\pi$ orbital becomes more
stable than the $2\si$ orbital at fields stronger than $ B_0$.

\subsection{Triplet states}

The total energy of the triplet state of the molecules calculated with
the two approaches agree well up to about $B = B_0$. When the field
strength is increased further, the basis set truncation error
increases again and approaches 1000 kcal/mol at $B=10~B_0$.

The \ce{H2} molecule did not behave well in the GTO calculations,
probably due to lack of basis functions needed for an accurate
description of the $1\pi$ orbital. The basis set truncation error is
11.5 kcal/mol at $B = B_0$.  For the other molecules, the basis set
truncation errors lie in the range of 0.2 -- 3.3 kcal/mol for $ 0 \leq
B \leq B_0 $.  The monotonically increasing basis set truncation
errors reach 30 kcal/mol at $B = 2~ B_0$.  At strong magnetic fields,
the errors reach hundreds of kcal/mol for \ce{H2} and \ce{HeH+}, and
1300 kcal/mol for \ce{BH}.

State crossings occur for many of the triplet states. In the \HelFEM{}
calculations, the triplet state of \ce{H2} exhibits crossings between
the $(1 \sigma_g 1\sigma_u)_\beta$ and $(1\sigma_g 1\pi_u)_\beta$
configurations at $B = 0.5~ B_0$. As was mentioned above, the
$(1\sigma 1\pi)$ state is not reproduced accurately by the GTO
calculation, which can tentatively be attributed to the insufficient
number of $p$ functions in the used basis set for hydrogen. This also
leads to a different state-crossing point compared to the numerical
calculation.

The calculations show that there is a state crossing at $B = 0.2~ B_0$
for the triplet state of \ce{BeH+} and at $B = 0.1~ B_0$ for the
triplet state of \ce{LiH}, where the $1\sigma_\alpha (1\sigma 2\sigma
1\pi)_\beta$ configuration becomes the ground state.

The triplet state of \ce{BH} was found to be challenging.  The
\London{} calculations yielded an energy of $-25.116456 E_h$ at $B=0$,
which is lower than the energy of $-25.115935 E_h$ obtained in the
fully numerical calculation.  Field-free calculations with a
development version of \Qchem{} 5.1.1\cite{Shao2015} yielded the same
energy as \London{}, whereas the total energy obtained with the
\Erkale{}\cite{erkale, Lehtola2012} and \PsiFour{}\cite{Parrish2017}
codes for a symmetry-abiding wave function is higher than the one
obtained in the fully numerical calculation.  Despite the significant
energy difference, the \PsiFour{} and \Qchem{} calculations converged
onto true local minima.  The different energies that were obtained
with the \London{}, \Erkale{}, \PsiFour{} and \Qchem{} programs
suggest that the calculations on the triplet state of \ce{BH}
converged to different solutions, even though the orbital
configurations supposedly match.  However, once the magnetic field is
switched on, the problem with multiple solutions vanishes. The same
solutions are obtained with the \London{} and \HelFEM{} codes for $B
\ge 0.001~B_0$.

The electron configuration of the triplet state of \ce{BH} is $(1
\sigma 2\sigma)_\alpha (1\sigma 2\sigma 3 \sigma 1\pi )_\beta$ in weak
fields.  The configuration of the $\alpha$ electrons changes from $
(1\si 2\si)_\al $ to $( 1\si 1\pi)_\al$ at $B = 1.5~ B_0$ according to
the \London{} calculations and at $B = 1.2~ B_0$ in the \HelFEM{}
calculations. The state with the $( 1\si 1\pi)_\al (1\si 2\si 3\si
1\pi)_\be $ configuration crosses the $( 1\si 1\pi)_\al (1\si 2\si
1\pi 2\pi)_\be $ state at $B = 3.3~ B_0$ in the fully numerical
calculations; this state crossing was not obtained in the GTO
calculations.

The triplet state of \ce{CH+} behaves in the same way as the triplet
state of \ce{BH}, except at $B = 0$. The weak field configuration is $
(1\si 2\si)_\al (1\si 2\si 3\si 1\pi)_\be $. In the GTO calculations,
a state crossing occurs at $B = 1.3~ B_0$, where the state with the $
(1\si 1\pi)_\al (1\si 2\si 3\si 1\pi)_\be $ becomes the ground
state. This state crossing occurs already at $B = 1.1~ B_0$ in the
fully numerical calculations.

\subsection{Quintet states}

The quintet states are well described by the GTO calculations in the
zero-field case, while the basis set truncation errors increase
significantly at finite field.  Large oscillations in the basis set
truncation error are observed for \ce{BH} and \ce{CH+}.

In the GTO calculations, the ground-state electron configuration of
the quintet state of \ce{LiH} is $(1\si 2\si 3\si 4\si)_\be$ at fields
weaker than $0.01~ B_0$; this configuration is an excited state in the
fully numerical calculations.  The $(1\si 2\si 3\si 1\pi)_\be$
configuration is the ground state of \ce{LiH} in the range of $ 0.01~
B_0 \leq B < 4.7~ B_0 $ in the GTO calculations, whereas a state
crossing from the $(1\si 2\si 3\si 1\pi)_\be$ configuration to $(1\si
2\si 1\pi 1\de)_\be$ occurs already at $B = 0.7~ B_0$ in the fully
numerical calculations. This state containing a $\delta$ orbital was
not found in the GTO calculations. Instead, a second state crossing is
seen in the GTO calculations at $B = 9.7~ B_0$, where the $(1\si 2\si
3\si 1\pi)_\be$ configuration becomes the ground state again. The
second state crossing is unexpected and is probably another artifact
due to the incomplete GTO basis set.

The quintet state of \ce{BeH+} has a $(1\si 2\si 3\si 1\pi)_\be$
configuration in the whole range of the investigated magnetic field
strengths in the GTO calculations. The ground state in the fully
numerical calculations, however, changes to $(1\si 2\si 1\pi
1\de)_\be$ at $B = 1.5~ B_0$, similarly to the case of LiH.

Many ground-state electron configurations were obtained for \ce{BH} in
the GTO calculations, whereas the numerical calculations yielded only
two ground-state configurations.  For $ 0 \leq B < 0.4~B_0 $, the
highest occupied $\beta$ orbital $({\pi}_{+})$ is a destabilized
$\pi_{+}$ orbital, whose orbital energy increases with increasing
strength of the magnetic field. The obtained electron configuration at
weak magnetic fields is $ 1\osa (1\si 2\si 3\si 1\pi 1{\pi}_{+})_\be $
in both the \HelFEM{} and \London{} calculations.  For magnetic fields
stronger than $B = 0.4~ B_0$, the \HelFEM{} calculations yielded a
ground-state configuration of $1\osa (1\si 2\si 3\si 1\pi 1\de)_\be $.
In the GTO calculations, the ground-state configuration in the range
of $B = [0.4, 0.7]~B_0$ is $1\osa (1\si 2\si 3\si 4\si 1\pi)_\be$,
whose $\pi$ orbital is stabilized when increasing the magnetic field
strength.  For $ 0.7~B_0 \leq B < 1.5~B_0 $, the obtained
configuration is $1\osa (1\si 2\si 3\si 1\pi 2\pi)_\be$.  The two
latter states are excited states in the fully numerical
calculations. The sharp peak at $B=0.4~B_0$ in the curve for the
quintet state in \figref{diffE} appears to originate from variations
in the basis set truncation error with respect to the strength of the
magnetic field.

For $ 1.5~B_0 \leq B \leq 4.0~B_0 $, a $1\osa (1\si 2\si 3\si 1\pi
1\de)_\be $ ground-state configuration is obtained in both the
\London{} and \HelFEM{} calculations, whereas at stronger fields the
ground-state configuration of the GTO calculations is again $1\osa
(1\si 2\si 3\si 4\si 1\pi)_\be $.  Due to the orbital-Zeeman term in
the Hamiltonian, it is unexpected that a high-angular-momentum orbital
with $m = -2$ becomes less stable than a $\sigma$ orbital with $m =
0$, when increasing the strength of the magnetic field. Thus, the
second state crossing in the GTO calculations is most likely an
artifact due to the employed GTO basis set.

The ground-state configuration of the quintet state of \ce{CH+} is
$1\osa (1\si 2\si 3\si 1\pi 1{\pi}_{+})_\be $ at weak fields.  The
first state crossing to $1\osa (1\si 2\si 3\si 1\pi 1\de)_\be $ occurs
at $B = 0.7~B_0 $ in the \HelFEM{} calculations and at $B = 1.2~B_0 $
in the \London{} calculations. The small peak at $B=0.7~B_0$ in the
curve for the quintet state in \figref{diffE} appears to be caused by
variations in the basis set quality with respect to the strength of
the external magnetic field.  A second state crossing is obtained in
the GTO calculations at $B = 6.5~B_0 $, where the $1\osa (1\si 2\si
3\si 4\si 1\pi)_\be $ configuration becomes the lowest one.  This is
again most likely an artifact due to the employed GTO basis set as in
the \ce{BH} case.

\subsection{Basis set study}

Basis set truncation errors were studied by performing calculations on
\ce{BeH+} as the representative molecule due to its rich state
diagram. Preliminary calculations showed that higher-angular-momentum
functions become more important in stronger magnetic fields, and that
deviations from the numerical reference energies became much smaller
when using larger basis sets.

The influence of higher-angular momentum functions was assessed by
performing calculations using uncontracted aug-cc-pVQZ and aug-cc-pV5Z
basis sets without $g$ or $h$ functions, in order to make the
calculations more tractable. This yields triple\nobreakdash-,
quadruple- and quintuple-$\zeta$ basis sets with the compositions of
$6s3p2d$, $7s4p3d2f$, and $9s5p4d3f$ for H, and $12s6p3d2f$,
$13s7p4d3f$, and $15s9p5d4f$ for Be, respectively. The results of the
calculations in these basis sets are shown in \tabref{bste}. The
basis-set truncation errors are larger for stronger magnetic fields,
as expected.

\begin{table}
  \caption{Basis set truncation errors in kcal/mol for \ce{BeH+} in
    uncontracted, gauge-including Cartesian Gaussian basis sets with
    primitives adopted from the aug-cc-pVTZ, aug-cc-pVQZ, and
    aug-cc-pV5Z basis sets without $g$ and $h$
    functions.\label{tab:bste}}
  \begin{tabularx}{\linewidth}{lcrrr}
    \hline
    Spin state & $B/B_0$ & triple-$\zeta$ & quadruple-$\zeta$ & quintuple-$\zeta$ \\
    \hline
    \hline
    singlet & 1 &  0.57 & 0.16 & 0.03 \\
            & 2 & 24.16 & 4.00 & 1.77 \\
    \\
    triplet & 1 &  1.03 & 0.34 & 0.14 \\
            & 2 & 10.37 & 2.01 & 1.00 \\
    \\
    quintet & 1 &   2.48  & 0.75  & 0.24 \\
            & 2 & 103.30 & 29.00 & 25.04 \\
    \hline
    \hline
  \end{tabularx}
\end{table}

The calculations show that the triple-$\zeta$ basis set describes the
singlet and triplet states of \ce{BeH+} better than its quintet
state. The basis is not large enough for the quintet state, as it
fails to reproduce the occupied $\delta$ orbital, which is, however,
found with the quadruple- and quintuple-$\zeta$ basis sets. This
manifests as the huge error -- over 100 kcal/mol -- at $B = 2
B_0$. But, the energy difference from the numerical reference value
for the same configuration is only 19.09 kcal/mol.

At strong magnetic fields, the ground state might have occupied
orbitals with larger $|m|$ quantum numbers. However, states with
occupied orbitals that have $|m|$ values larger than 2 ($\delta$
orbitals) have not been investigated in the present work. The
benchmark study suggests that rather accurate energies can be obtained
for molecules in magnetic fields stronger than $B= 1~B_0$, provided
that a GTO basis set providing sufficient coverage for the occupied
orbitals in its range of exponents is used. For instance, a proper
description of the $\delta$ orbital in the quintet state of \ce{BeH+}
requires a large set of $d$ functions, as well as
higher-angular-momentum functions to describe the polarization of the
$\delta$ orbital.

\section{Summary and Conclusion}
\label{sec:summary}

We have performed fully numerical Hartree--Fock (HF) calculations on
diatomic molecules exposed to an external magnetic field along the
molecular axis with the \HelFEM{} program. The main aim of the study
was to assess the basis set truncation errors in unrestricted HF
energies obtained using large gauge-including basis sets of
uncontracted Gaussian-type orbitals (GTO). Although calculations at
the density functional theory (DFT) level can also be performed with
\HelFEM{}, for simplicity we chose the HF level for the present study,
since HF and DFT calculations have similar basis set requirements.

We performed calculations on a few low-lying states of \ce{H2},
\ce{HeH+}, \ce{LiH}, \ce{BeH+}, \ce{BH}, and \ce{CH+} as a function of
the magnetic field strength in the range of $B=[0,10]~B_0$, using the
recently published \HelFEM{} program\cite{Lehtola2019a, Lehtola2019b,
  HelFEM} that has been extended to calculations on atoms and diatomic
molecules in finite magnetic fields in the present work. Due to state
crossings, several electronic states were studied as a function of the
magnetic field strength. As the aim was to study basis set effects,
fixed internuclear distances were employed, even though bond lengths
are known to decrease significantly when increasing the strength of
the magnetic field.
 
In the present work, we compared energies calculated using the
numerical approach implemented in the \HelFEM{} code with values
calculated with the \London{} program, employing London-
a.k.a. gauge-including atomic orbitals (GIAO) formed from
commonly-used primitive Gaussian basis sets.  As the \HelFEM{}
calculations employ local basis sets close to the complete basis set
limit, gauge factors such as those used in \London{} are not necessary
in the fully numerical approach. The energies obtained in the fully
numerical UHF calculations can be used in other studies as reference
data.

The calculations show that the Cartesian forms of the uncontracted
aug-cc-pVTZ basis sets\cite{Dunning1989, Woon1995} have basis set
truncation errors of the order of 1 kcal/mol in the absence of the
external magnetic field, whereas the truncation errors of the
gauge-including basis become significantly larger when the strength of
the magnetic field is increased.  The largest basis set truncation
error at $B = 10\ B_0$ is more than 1000 kcal/mol. The calculations
also show oscillations in the energy as a function of the magnetic
field, which can be possibly explained by the basis set truncation
error being smaller for some magnetic field strengths, for which the
exponents happen to be closer to optimal values than at other field
strengths.

Calculations on \ce{BeH+} with larger basis sets showed that the basis
set truncation errors can be made significantly smaller by employing
basis sets with more high-angular-momentum functions. The problem of
determining optimal atomic-orbital basis sets for finite fields could
be investigated in future work.

The comparison of the energies calculated using the two approaches
revealed lots of technical and formal difficulties. The ground-state
electron configurations change when increasing the strength of the
external magnetic field. When symmetry restrictions are not employed,
optimization algorithms may not always be able to find the
lowest-lying configuration, which leads to discontinuities in the
energy as a function of the magnetic field strength. The
discontinuities can, however, be resolved by tracking the various
states by reading in converged orbitals for the wanted configuration
at a different value of $B$.

We also discovered that in some cases the lowest state corresponds to
a symmetry-broken solution. Symmetry-broken solutions were a problem
only in the GTO calculations, as symmetry restrictions are not
supported in the \London{} program, whereas in the fully numerical \HelFEM{}
program the symmetry of the occupied orbitals were explicitly
enforced. \HelFEM{} also supports calculations of states with broken
symmetry. However, locating symmetry-broken solutions in the extended
basis set is not trivial.\cite{Lehtola2019a, Lehtola2019b}

\section*{Acknowledgment}

We thank Trygve Helgaker and Erik Tellgren for a copy of the \London{}
code, and Stella Stopkowicz and Florian Hampe for their tool for
determining orbital symmetry. This work was supported by The Academy
of Finland through projects 311149 and 314821 as well as by the Magnus
Ehrnrooth Foundation. The authors acknowledge CSC -- the Finnish IT
Center for Science as well as the Finnish Grid and Cloud
Infrastructure (persistent identifier
urn:nbn:fi:research-infras-2016072533) for computer time.

\bibliographystyle{tfo}
\bibliography{citations,london}

\newpage
\section*{Supporting Information}

\setcounter{table}{0}
\renewcommand{\thetable}{S\arabic{table}}
\setcounter{figure}{0}
\renewcommand{\thefigure}{S\arabic{figure}}

\begin{figure}
  \centering
  \subfloat[][GTO]{\includegraphics[width=.49\textwidth]{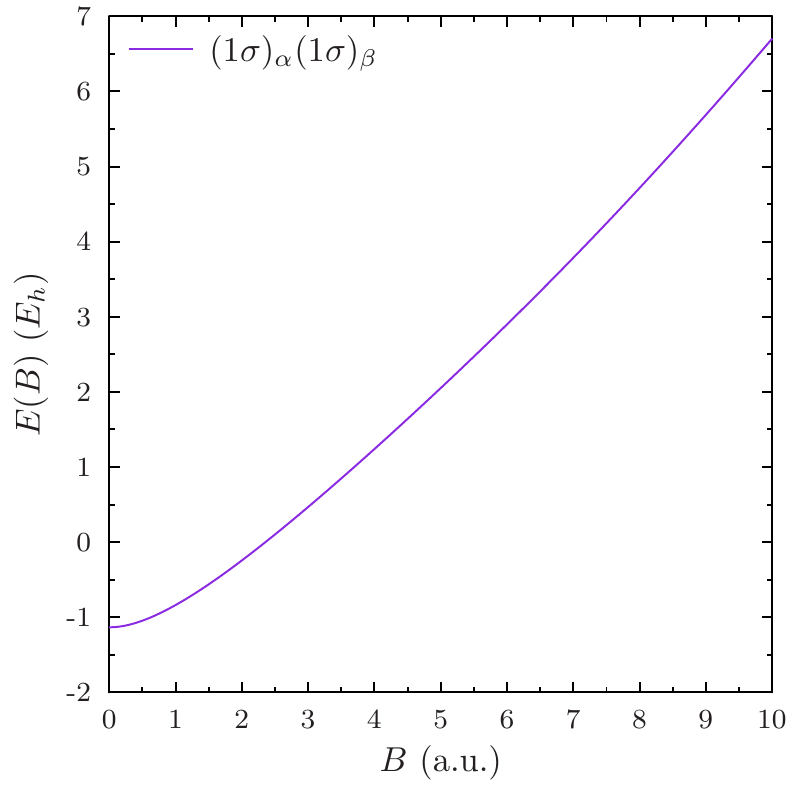}}
  \subfloat[][FEM]{\includegraphics[width=.49\textwidth]{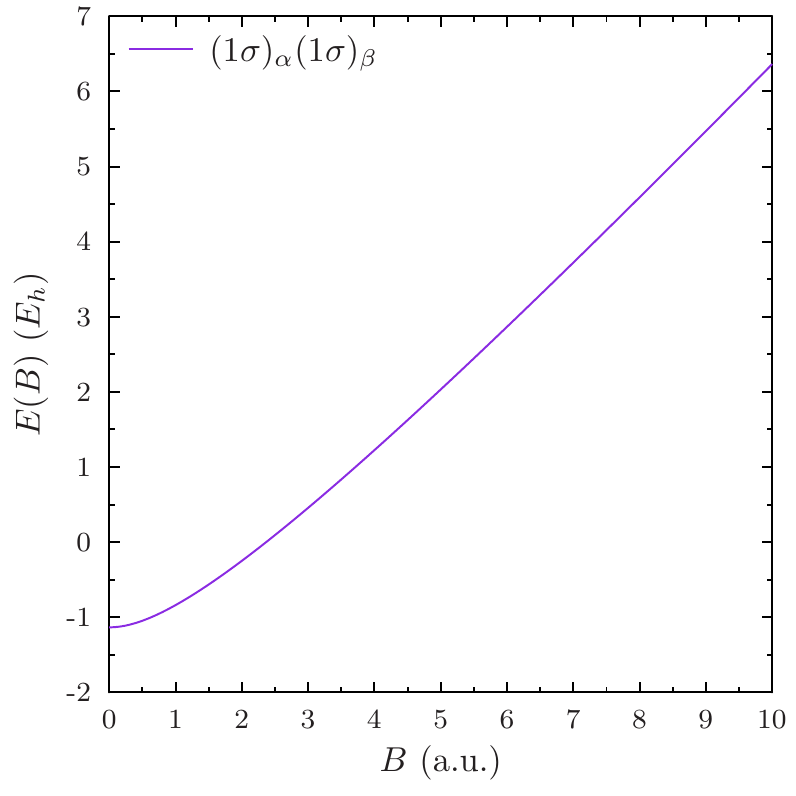}}
  \caption{\ce{H2} singlet.\label{fig:HH-1}}
\end{figure}

\begin{figure}
  \centering
  \subfloat[][GTO]{\includegraphics[width=.49\textwidth]{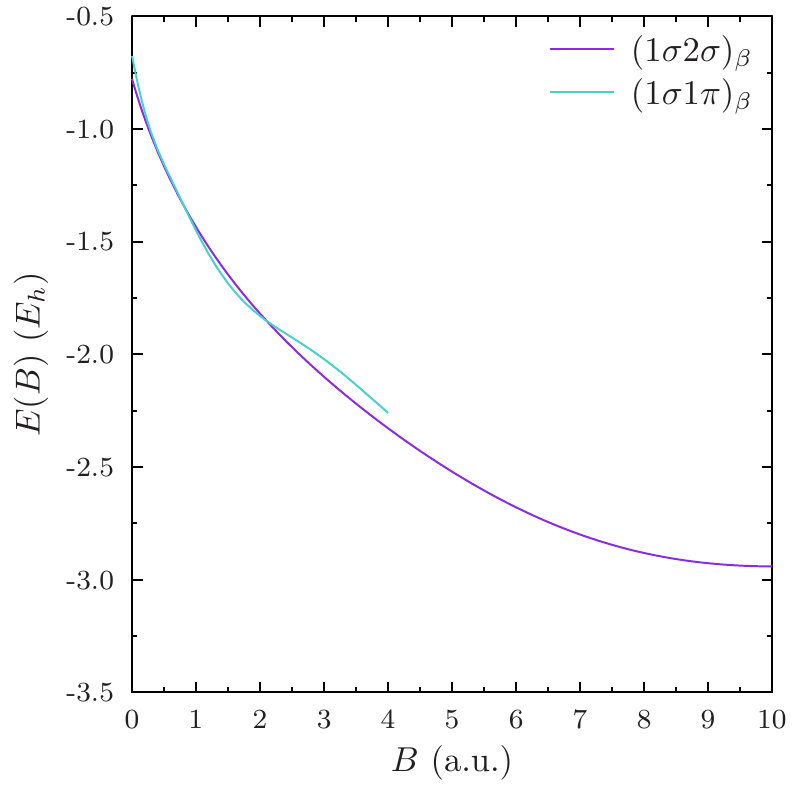}}
  \subfloat[][FEM]{\includegraphics[width=.49\textwidth]{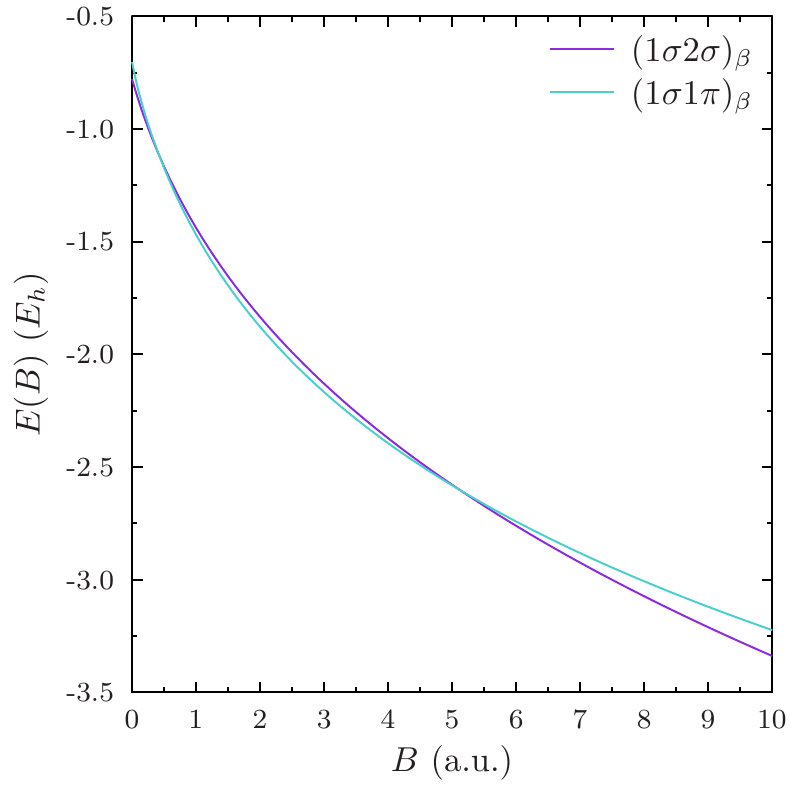}}
  \caption{\ce{H2} triplet.\label{fig:HH-3}}
\end{figure}

\begin{figure}
  \centering
  \subfloat[][GTO]{\includegraphics[width=.49\textwidth]{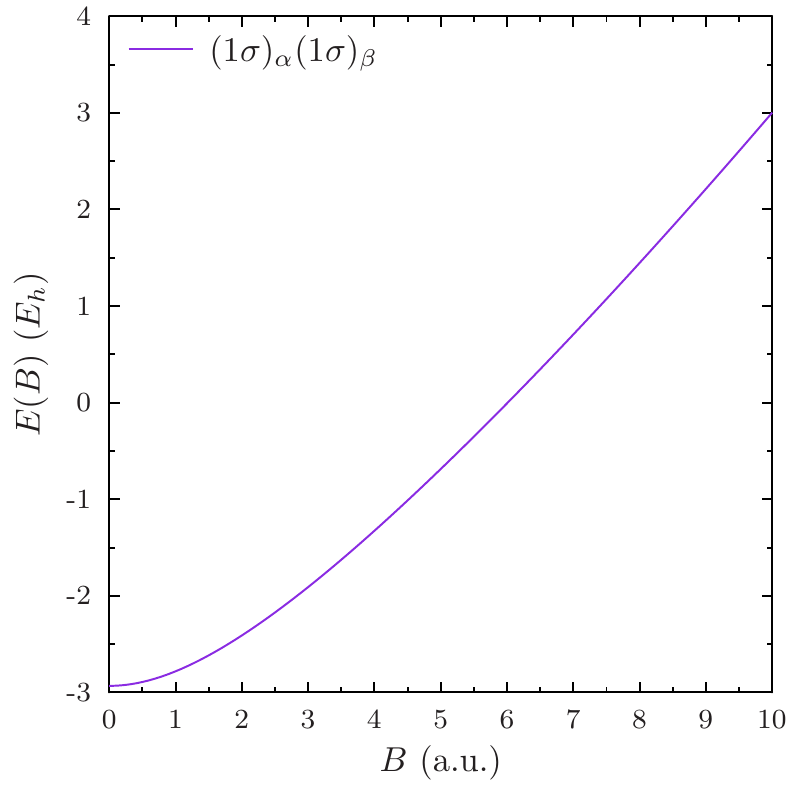}}
  \subfloat[][FEM]{\includegraphics[width=.49\textwidth]{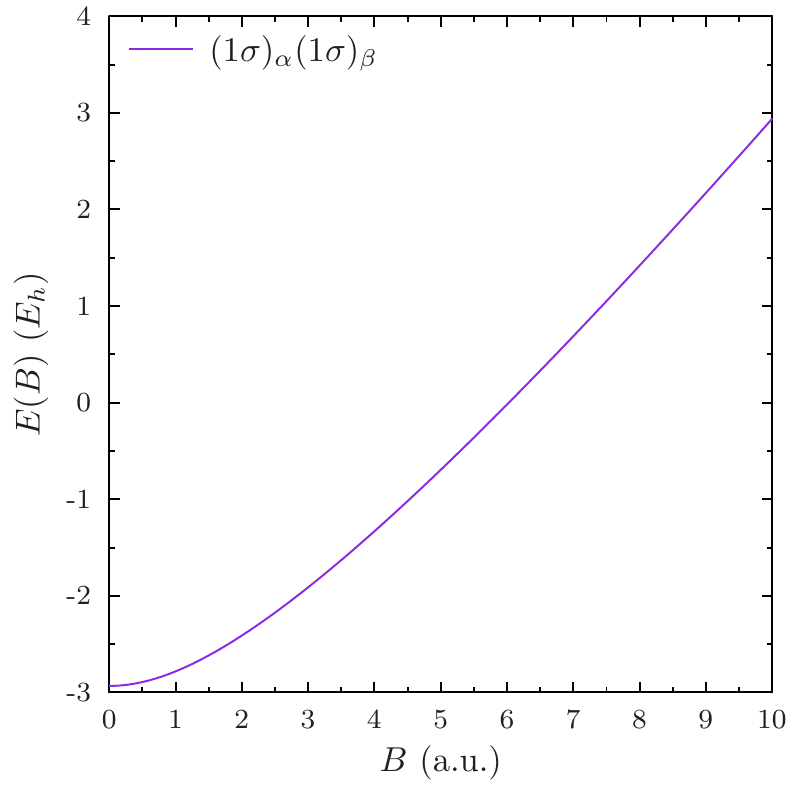}}
  \caption{\ce{HeH+} singlet.\label{fig:HeH+-1}}
\end{figure}

\begin{figure}
  \centering
  \subfloat[][GTO]{\includegraphics[width=.49\textwidth]{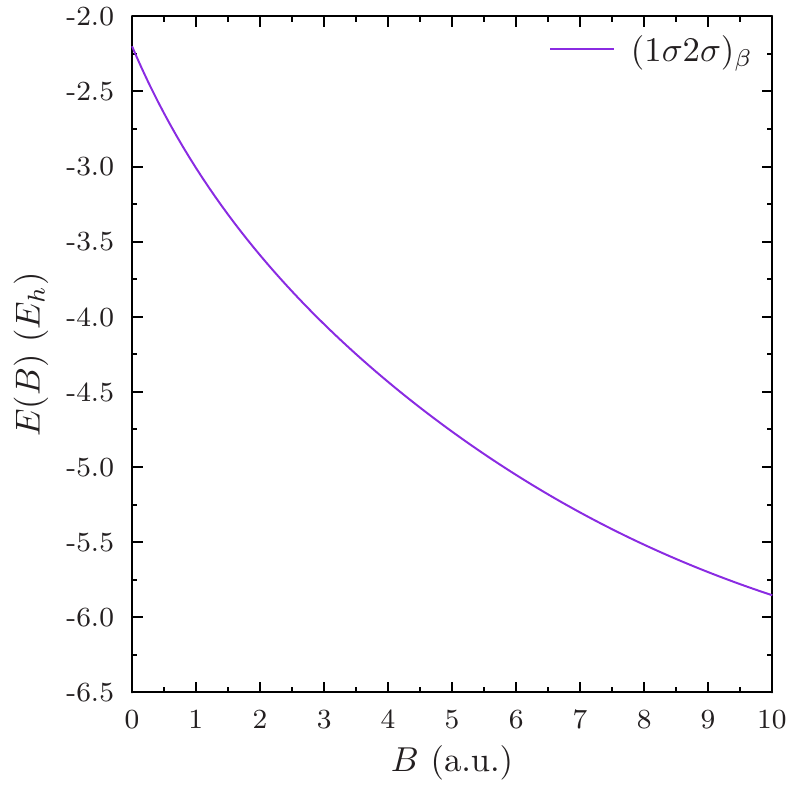}}
  \subfloat[][FEM]{\includegraphics[width=.49\textwidth]{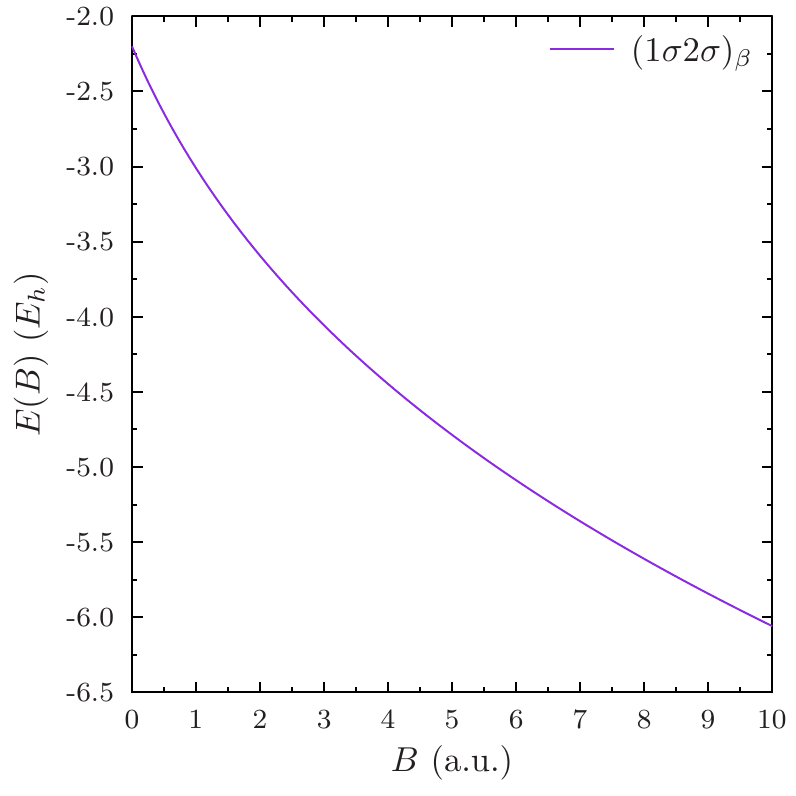}}
  \caption{\ce{HeH+} triplet.\label{fig:HeH+-3}}
\end{figure}

\begin{figure}
  \centering
  \subfloat[][GTO]{\includegraphics[width=.49\textwidth]{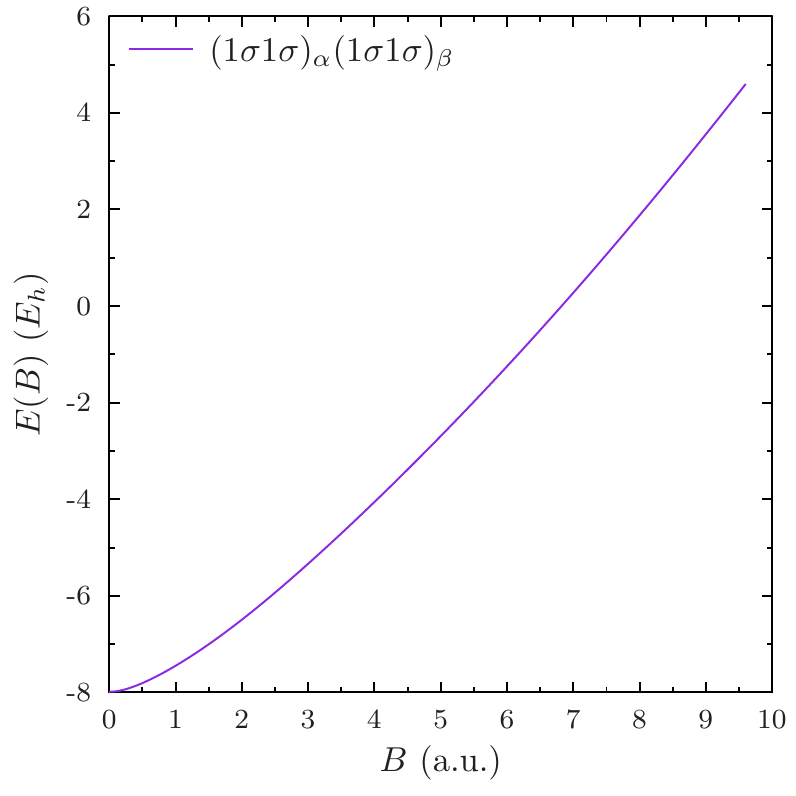}}
  \subfloat[][FEM]{\includegraphics[width=.49\textwidth]{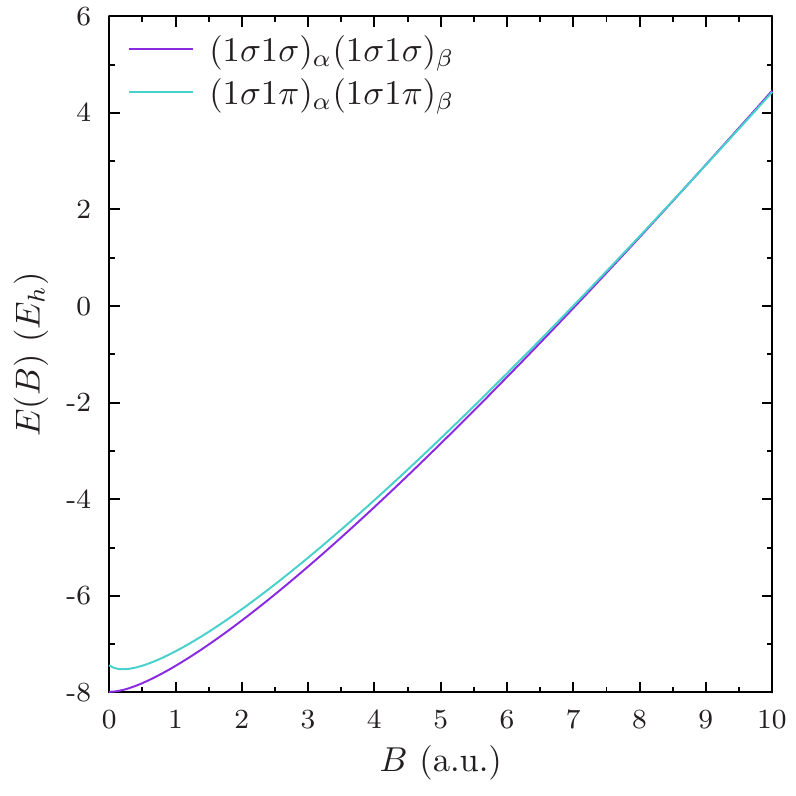}}
  \caption{\ce{LiH} singlet.\label{fig:LiH-1}}
\end{figure}

\begin{figure}
  \centering
  \subfloat[][GTO]{\includegraphics[width=.49\textwidth]{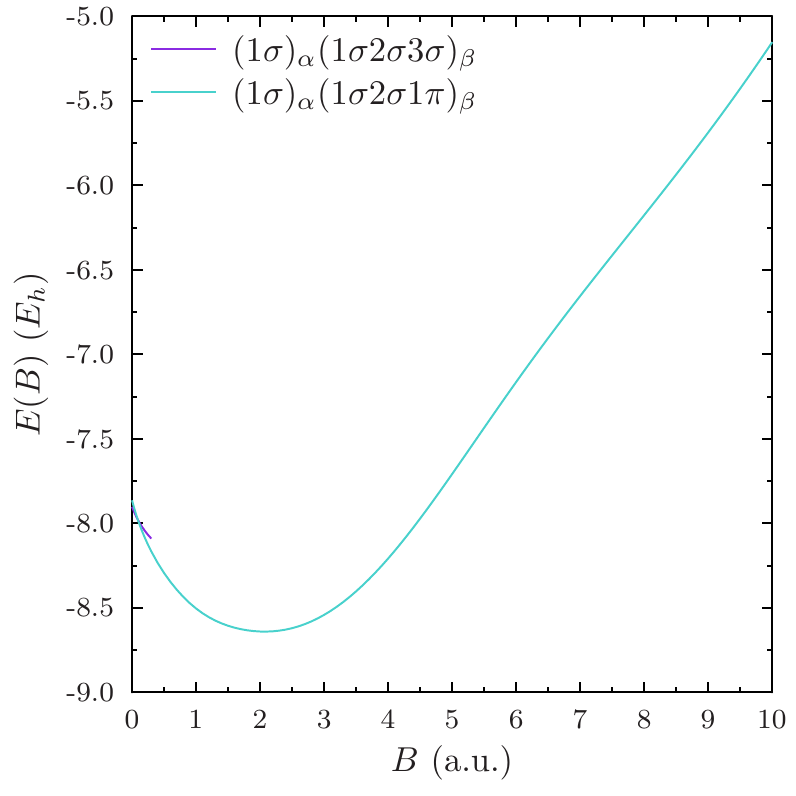}
    \includegraphics[width=.49\textwidth]{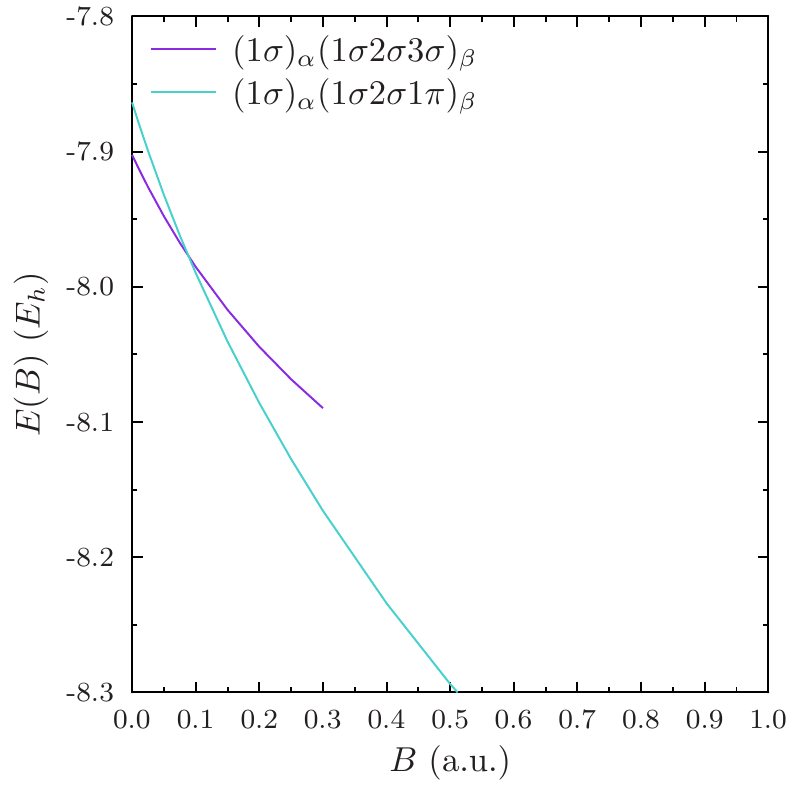}}

  \subfloat[][FEM]{\includegraphics[width=.49\textwidth]{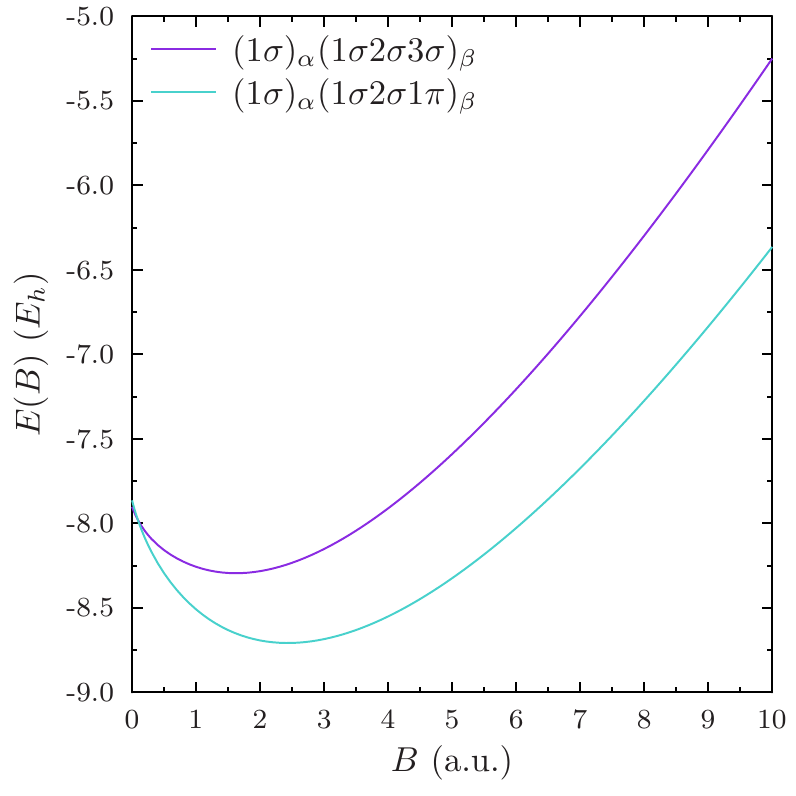}
    \includegraphics[width=.49\textwidth]{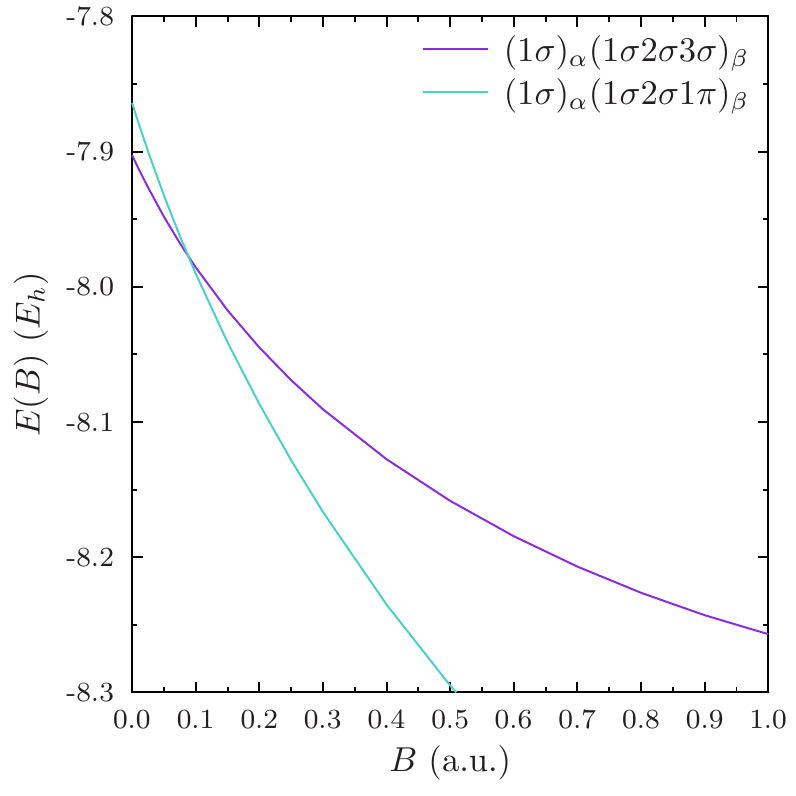}}
  \caption{\ce{LiH} triplet.\label{fig:LiH-3}}
\end{figure}

\begin{figure}
  \centering
  \subfloat[][GTO]{\includegraphics[width=.49\textwidth]{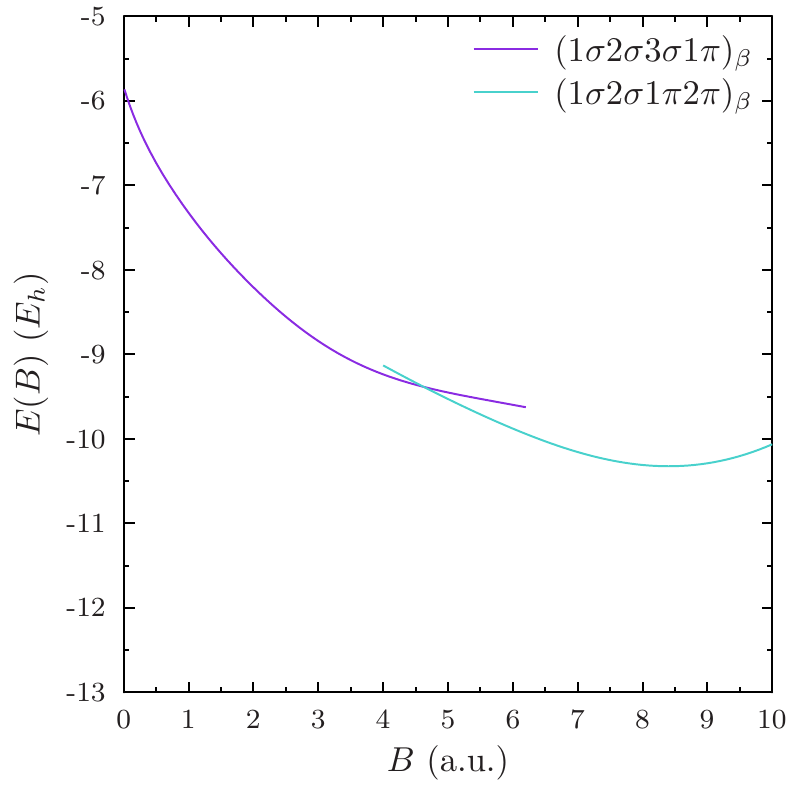}}
  \subfloat[][FEM]{\includegraphics[width=.49\textwidth]{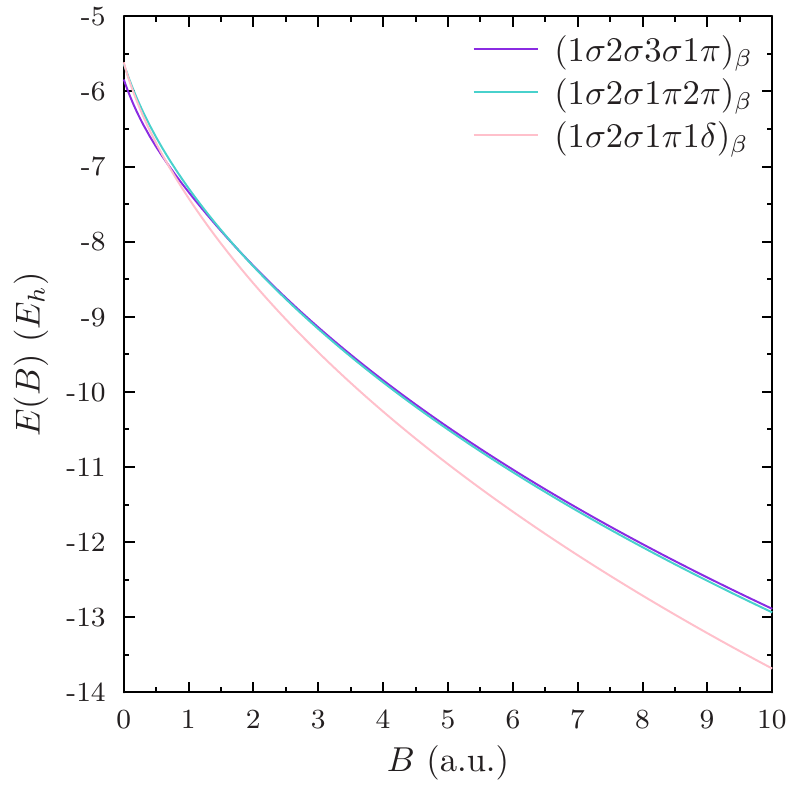}}
  \caption{\ce{LiH} quintet.\label{fig:LiH-5}}
\end{figure}

\begin{figure}
  \centering
  \subfloat[][GTO]{\includegraphics[width=.49\textwidth]{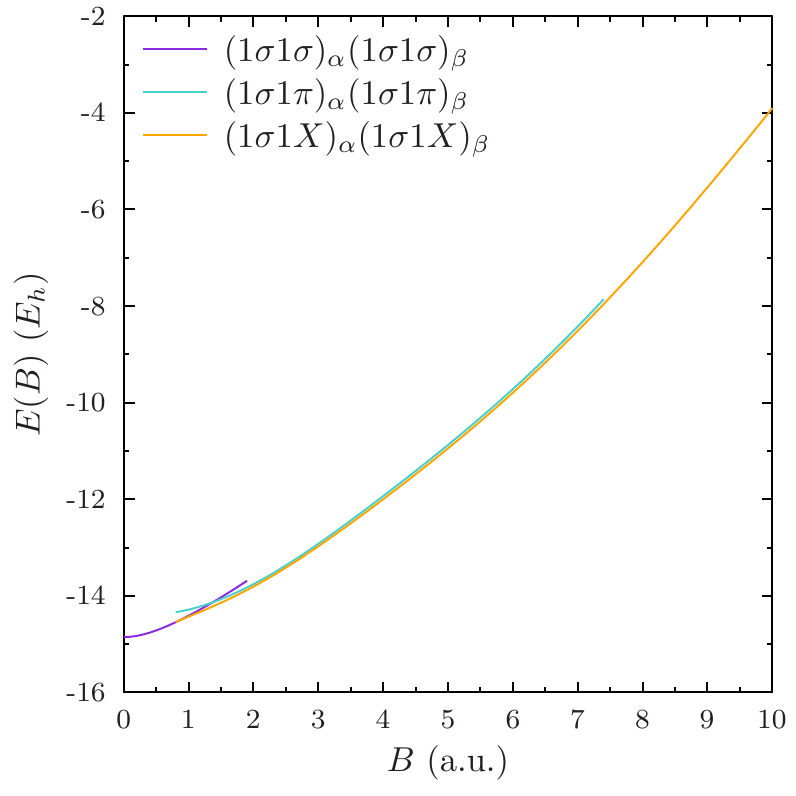}
    \includegraphics[width=.49\textwidth]{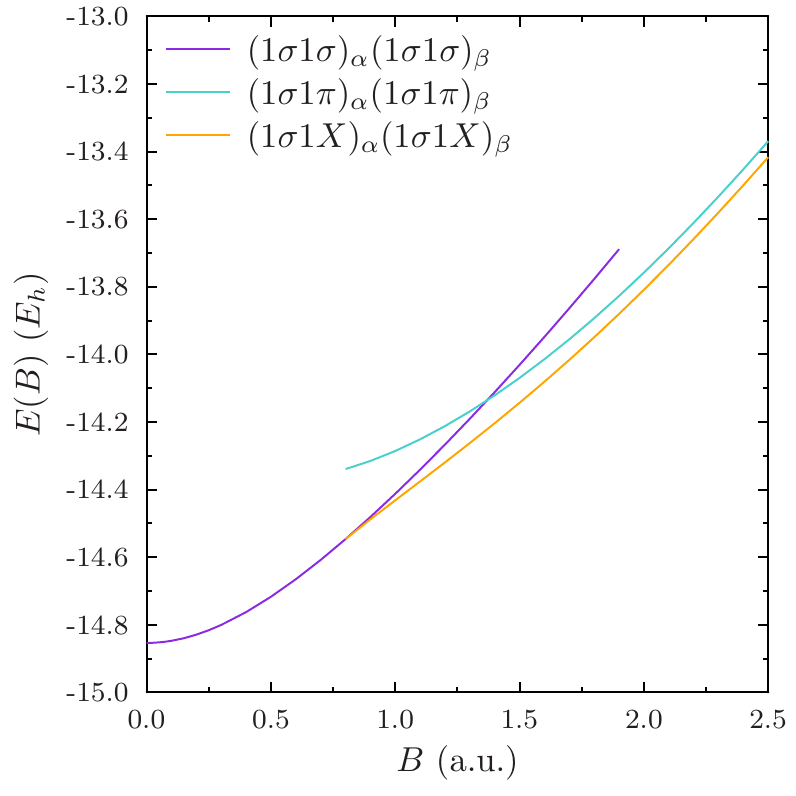}}

  \subfloat[][FEM]{\includegraphics[width=.49\textwidth]{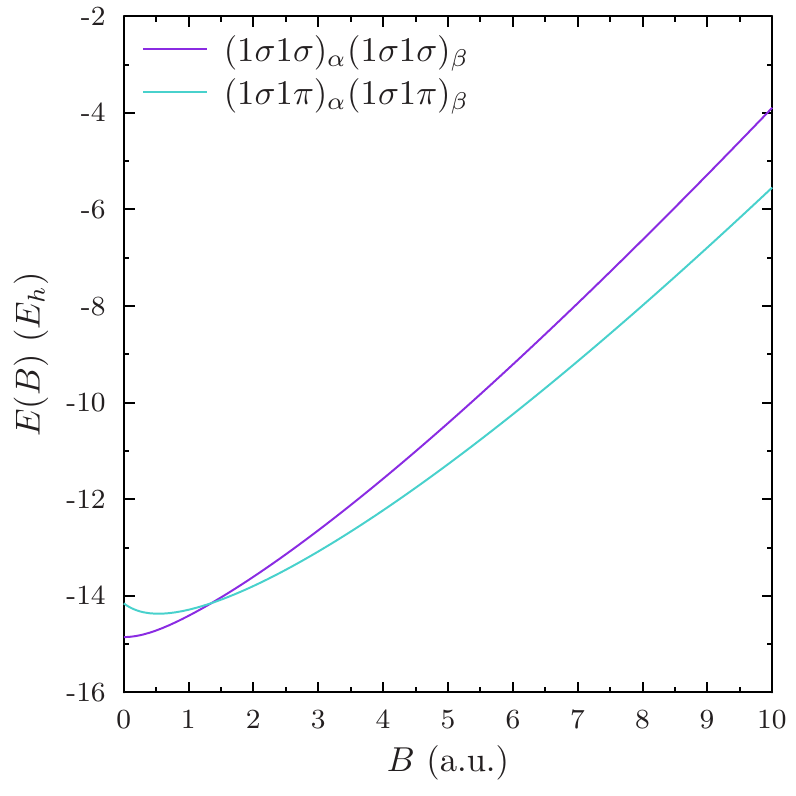}}
  \caption{\ce{BeH+} singlet.\label{fig:BeH+-1}}
\end{figure}

\begin{figure}
  \centering
  \subfloat[][GTO]{\includegraphics[width=.49\textwidth]{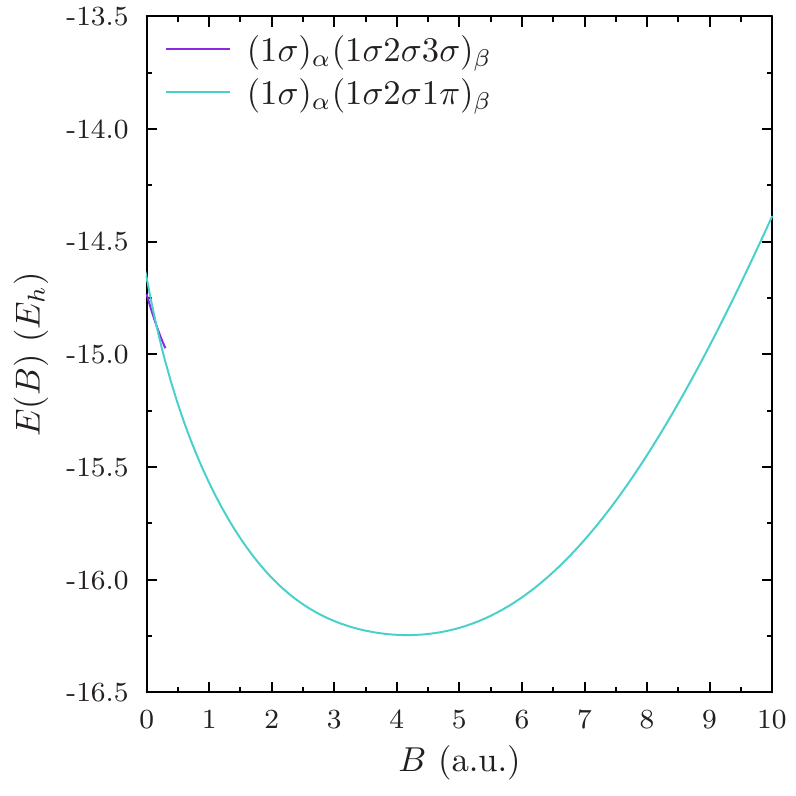} 
     \includegraphics[width=.49\textwidth]{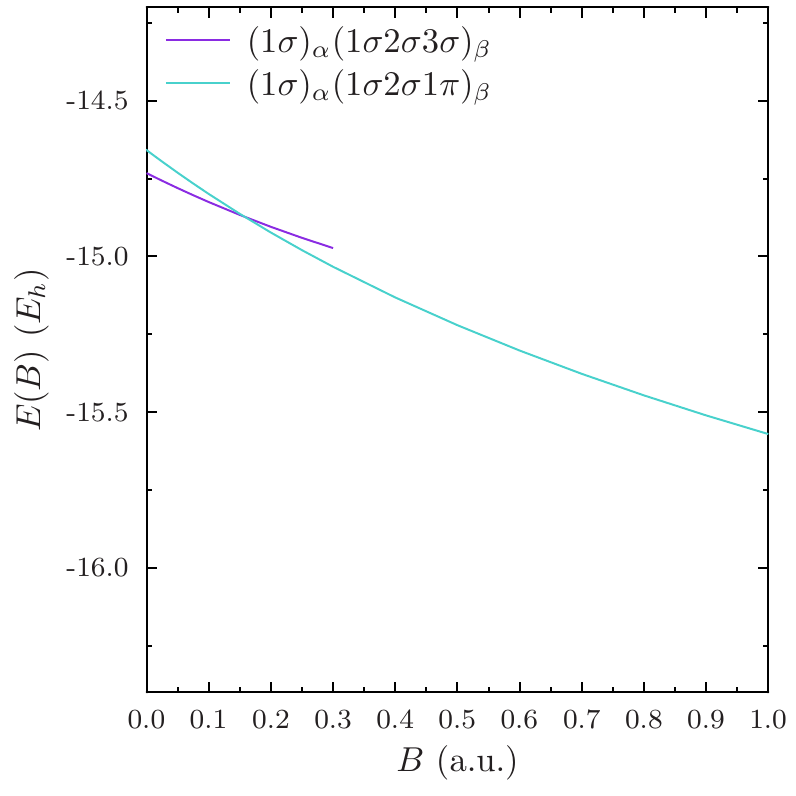}}

  \subfloat[][FEM]{\includegraphics[width=.49\textwidth]{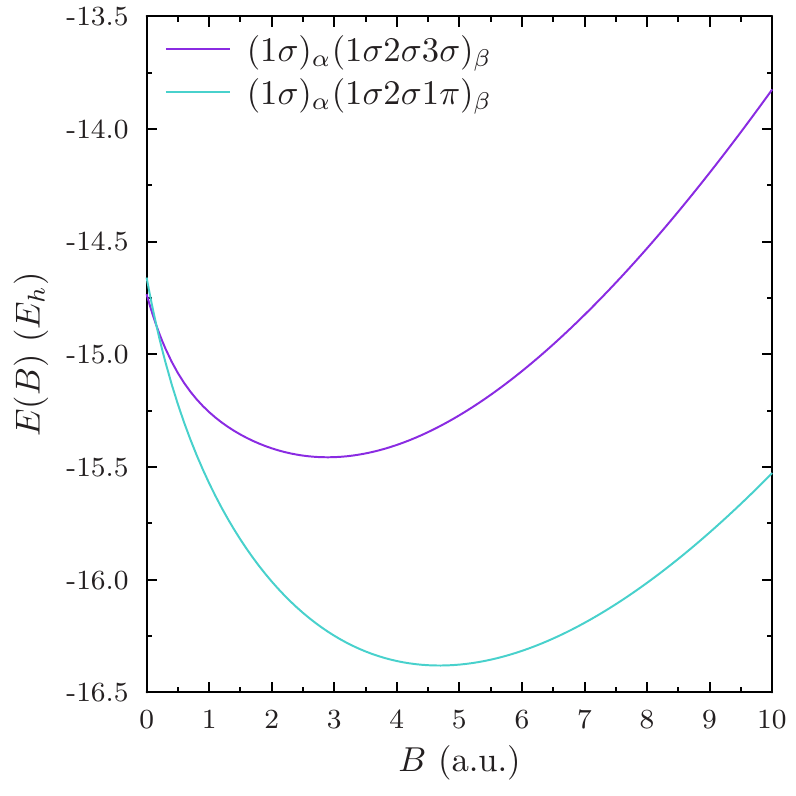}
    \includegraphics[width=.49\textwidth]{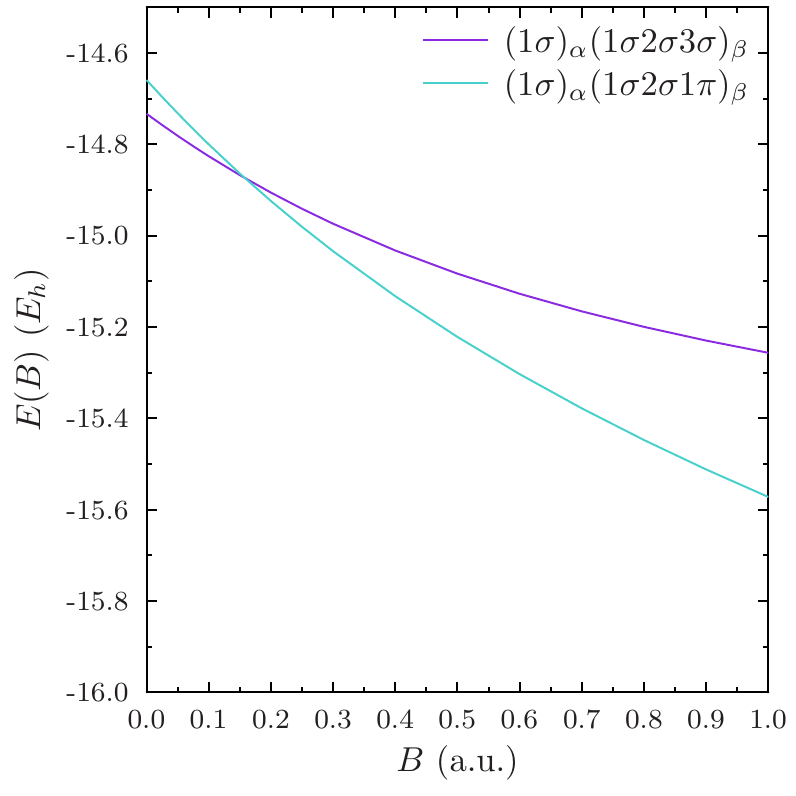}}
  \caption{\ce{BeH+} triplet.\label{fig:BeH+-3}}
\end{figure}

\begin{figure}
  \centering
  \subfloat[][GTO]{\includegraphics[width=.49\textwidth]{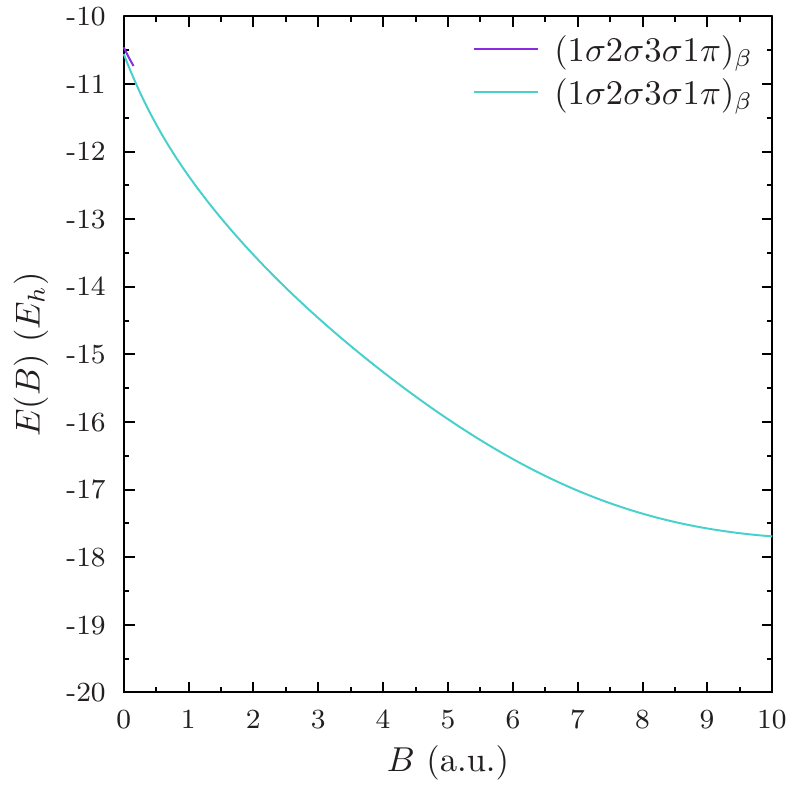}}
  \subfloat[][FEM]{\includegraphics[width=.49\textwidth]{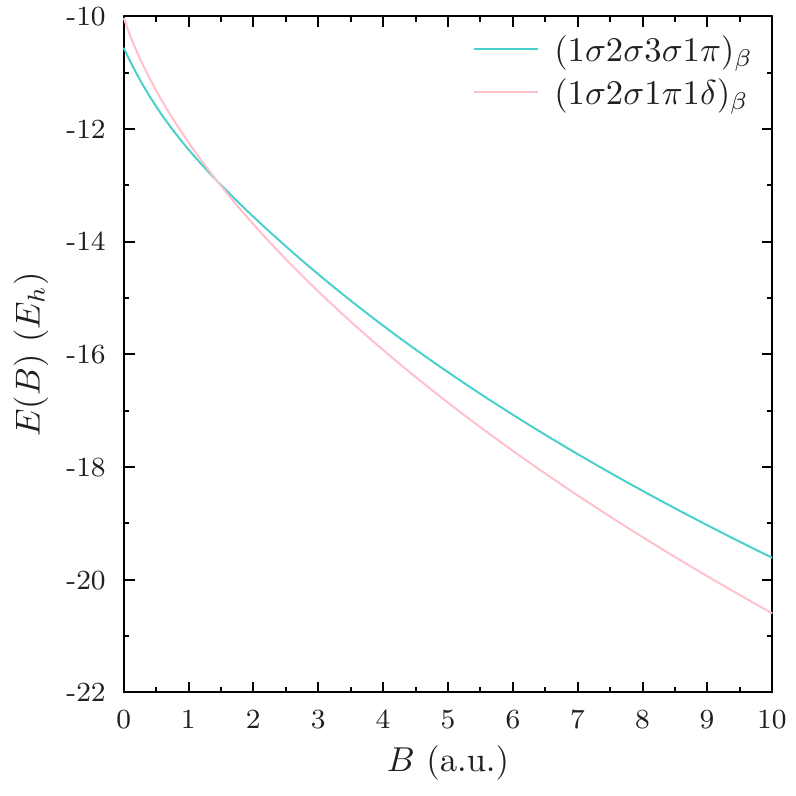}}
  \caption{\ce{BeH+} quintet.\label{fig:BeH+-5}}
\end{figure}

\begin{figure}
  \centering
  \subfloat[][GTO]{\includegraphics[width=.49\textwidth]{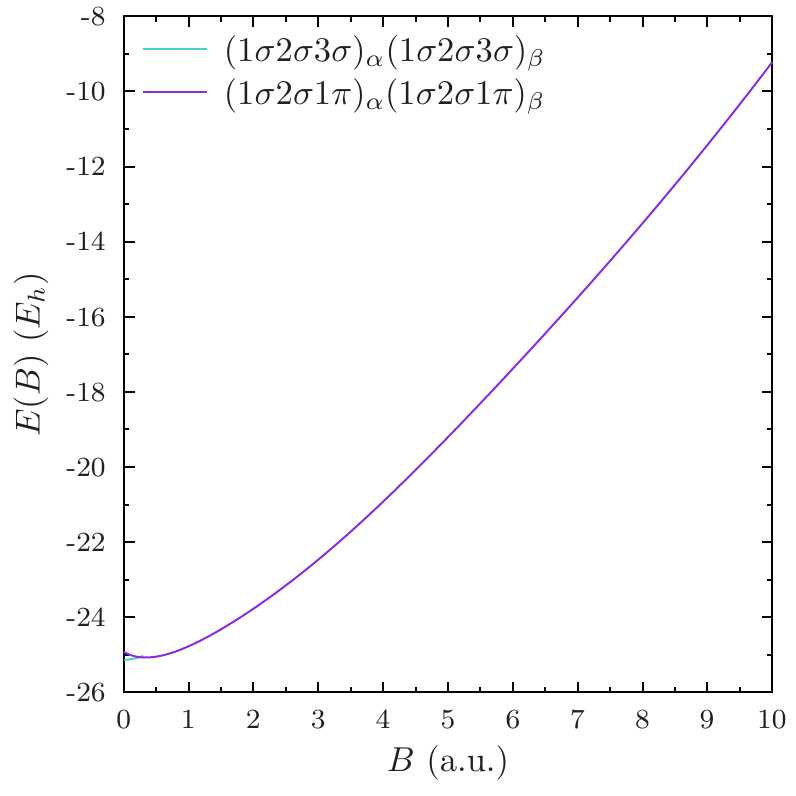}
    \includegraphics[width=.49\textwidth]{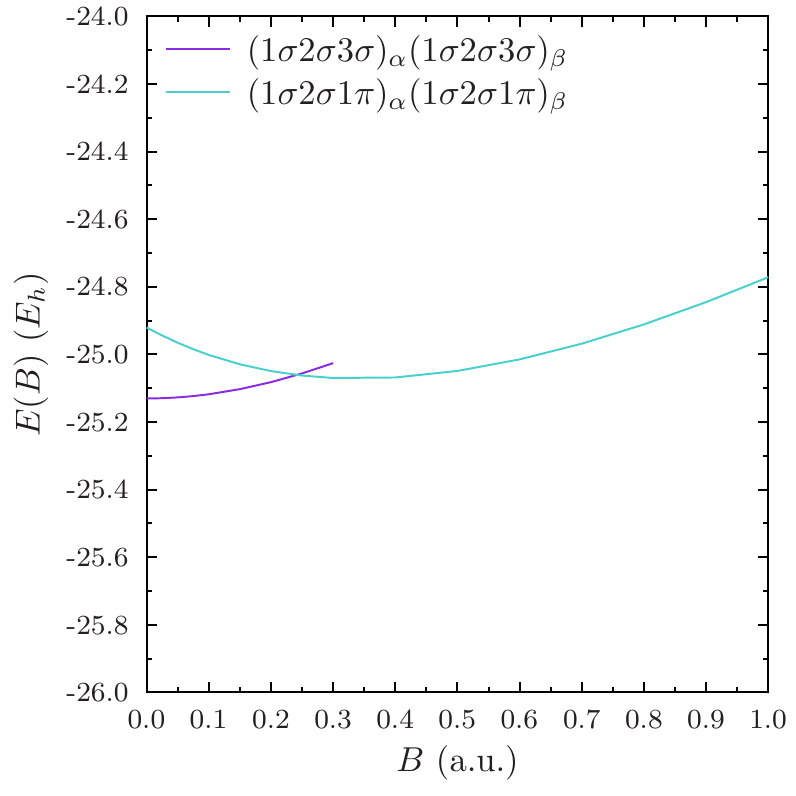}}

  \subfloat[][FEM]{\includegraphics[width=.49\textwidth]{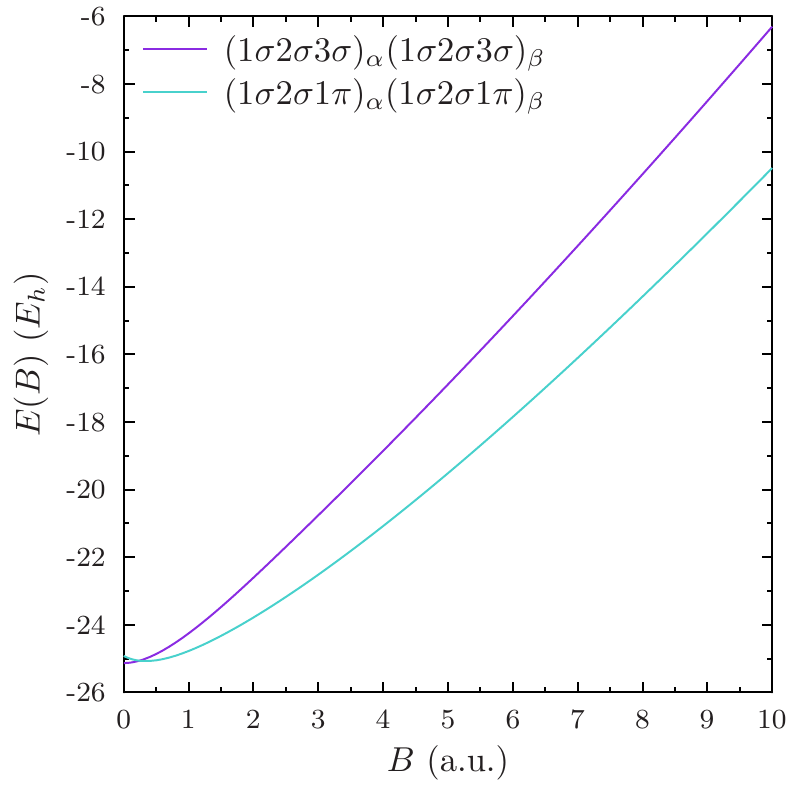}
    \includegraphics[width=.49\textwidth]{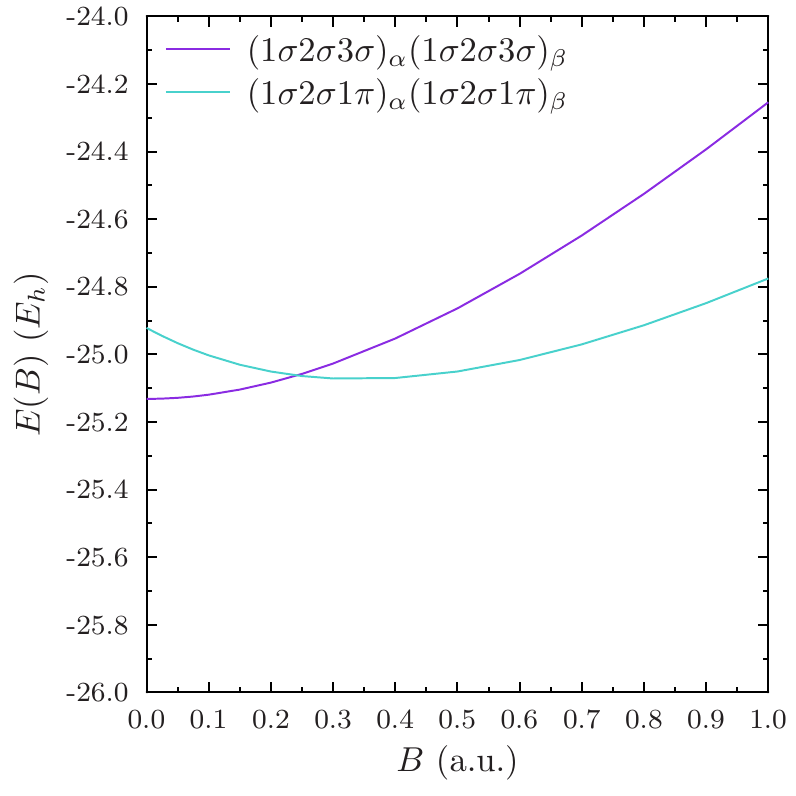}}
  \caption{\ce{BH} singlet.\label{fig:BH-1}}
\end{figure}

\begin{figure}
  \centering
  \subfloat[][GTO]{\includegraphics[width=.49\textwidth]{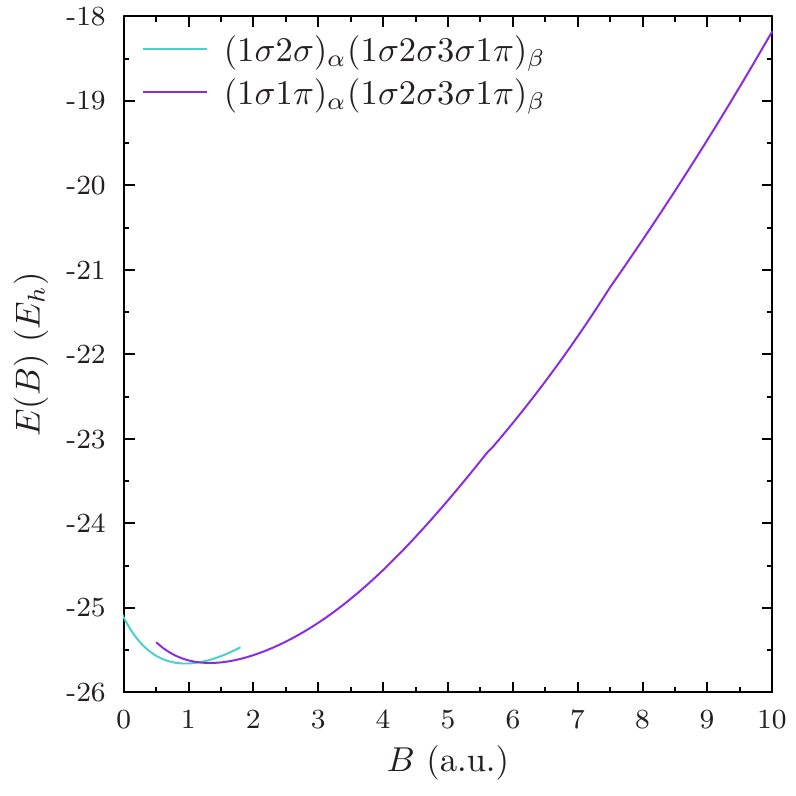}}
  \subfloat[][FEM]{\includegraphics[width=.49\textwidth]{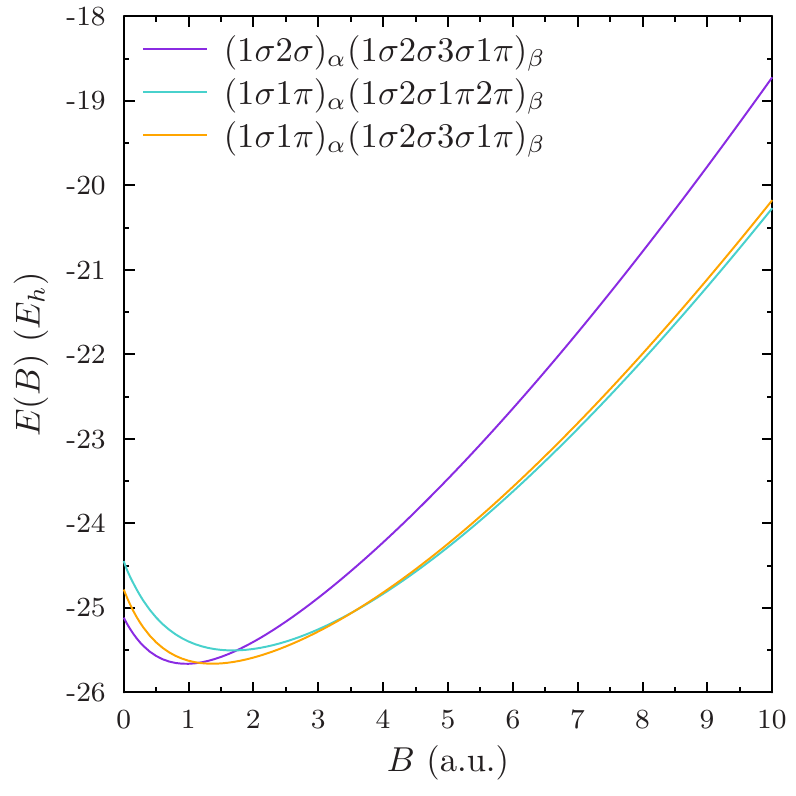}}
  \caption{\ce{BH} triplet.\label{fig:BH-3}}
\end{figure}

\begin{figure}
  \centering
  \subfloat[][GTO]{\includegraphics[width=.49\textwidth]{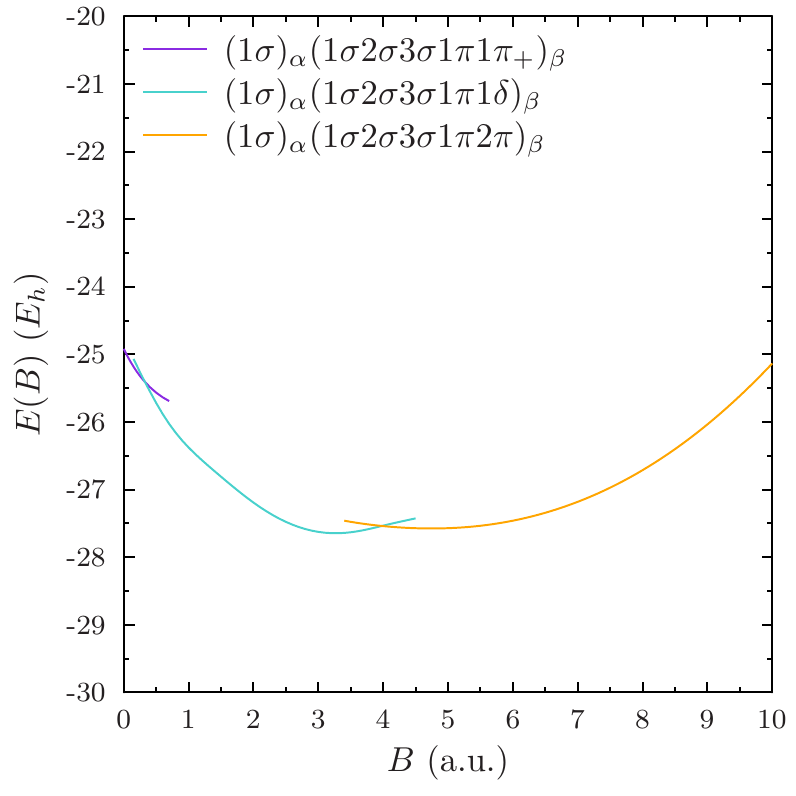}
    \includegraphics[width=.49\textwidth]{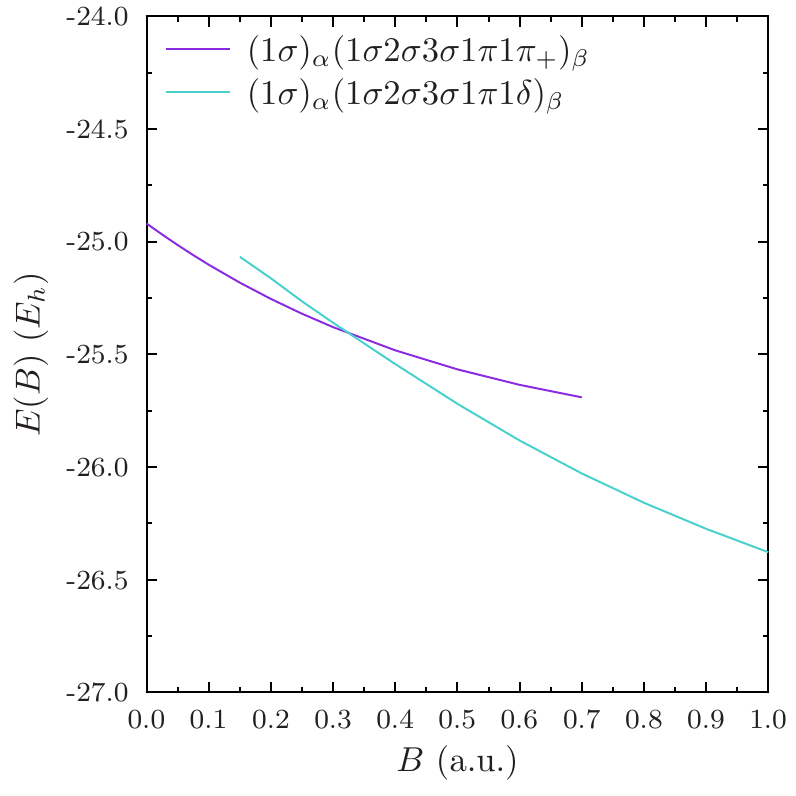}}

  \subfloat[][FEM]{\includegraphics[width=.49\textwidth]{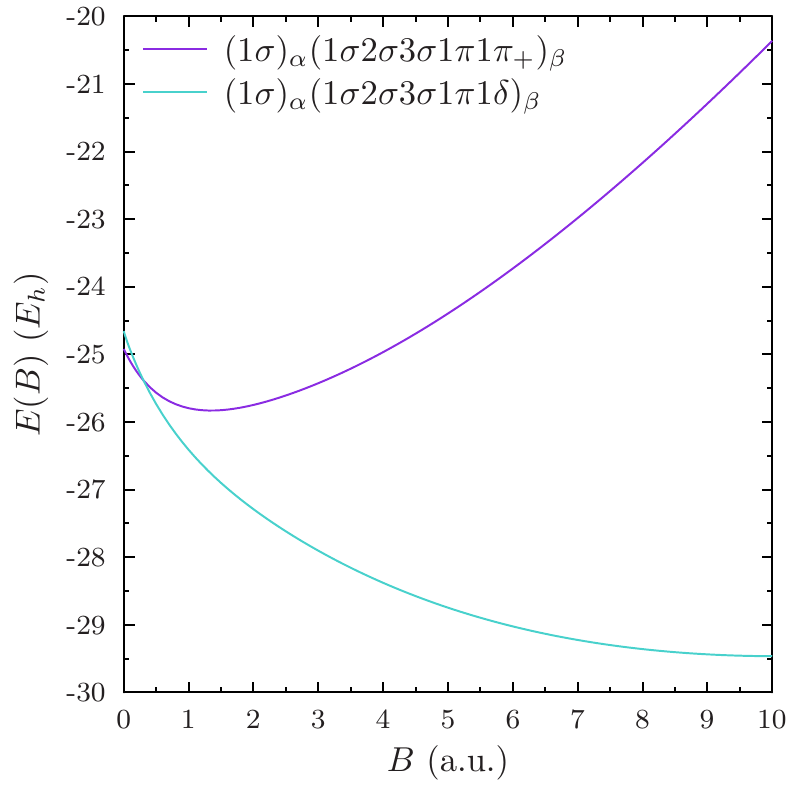}
    \includegraphics[width=.49\textwidth]{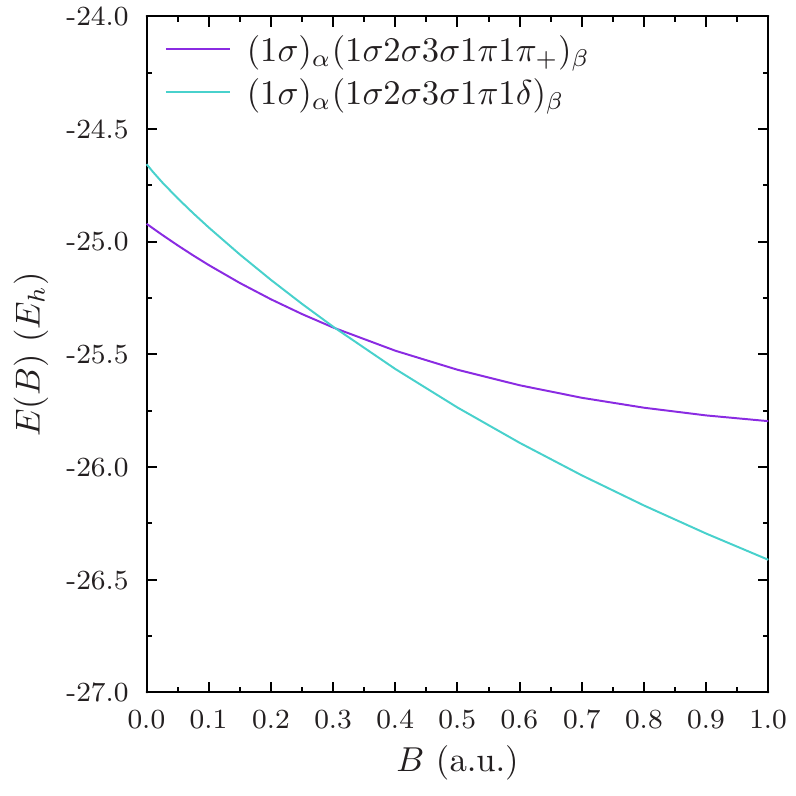}}
  \caption{\ce{BH} quintet.\label{fig:BH-5}}
\end{figure}

\begin{figure}
  \centering
  \subfloat[][GTO]{\includegraphics[width=.49\textwidth]{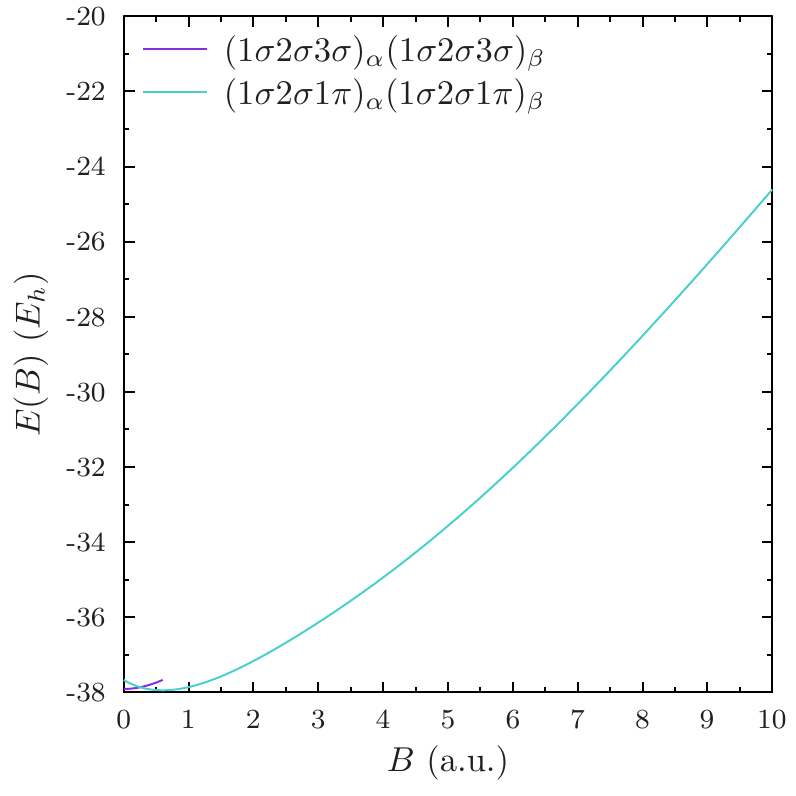}
    \includegraphics[width=.49\textwidth]{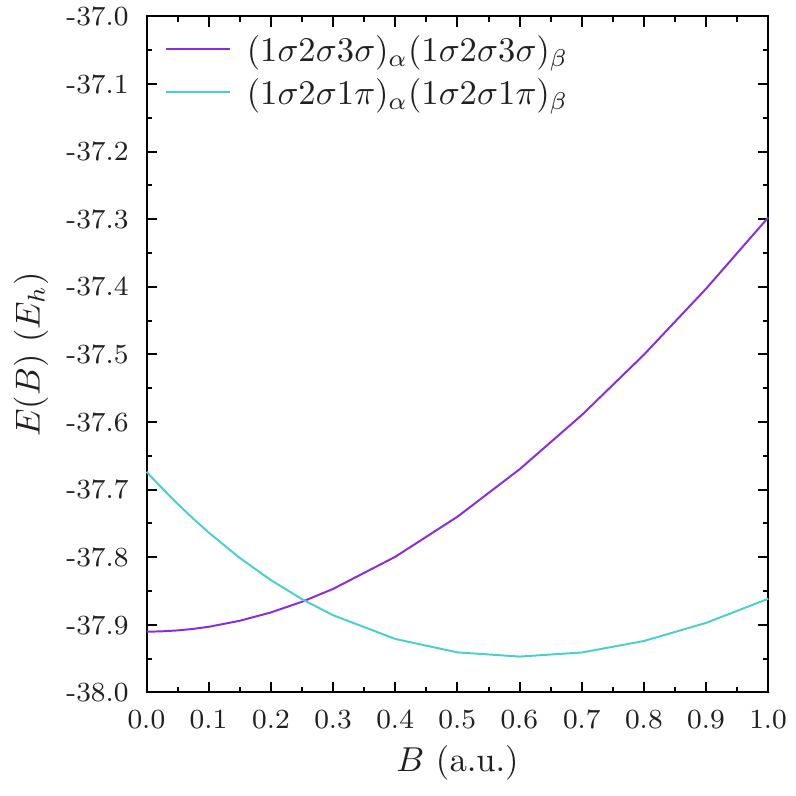}}

  \subfloat[][FEM]{\includegraphics[width=.49\textwidth]{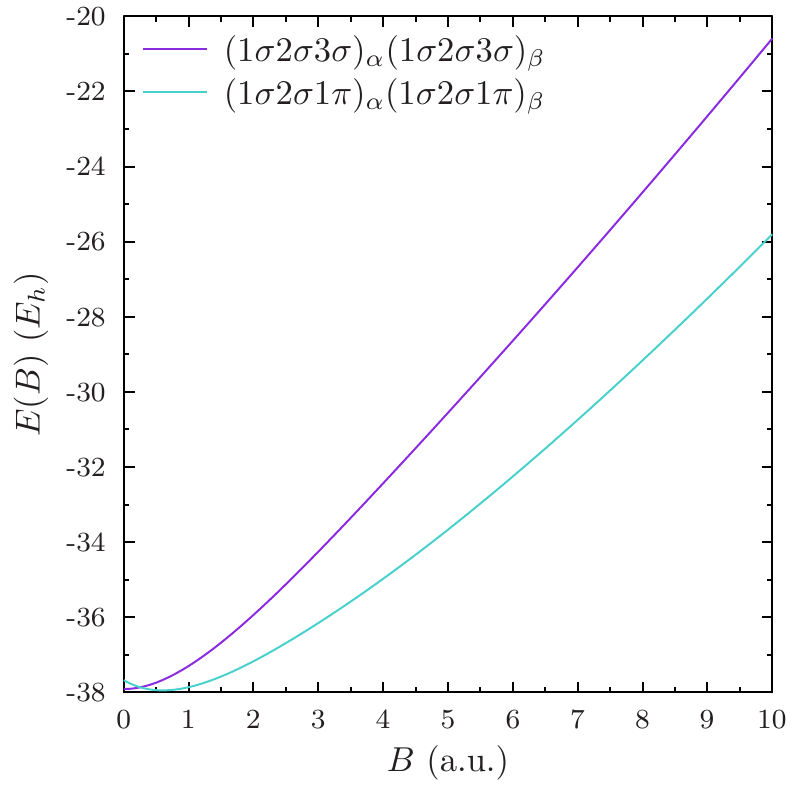}
    \includegraphics[width=.49\textwidth]{CH+_1_smallB}}
  \caption{\ce{CH+} singlet.\label{fig:CH+-1}}
\end{figure}

\begin{figure}
  \centering
  \subfloat[][GTO]{\includegraphics[width=.49\textwidth]{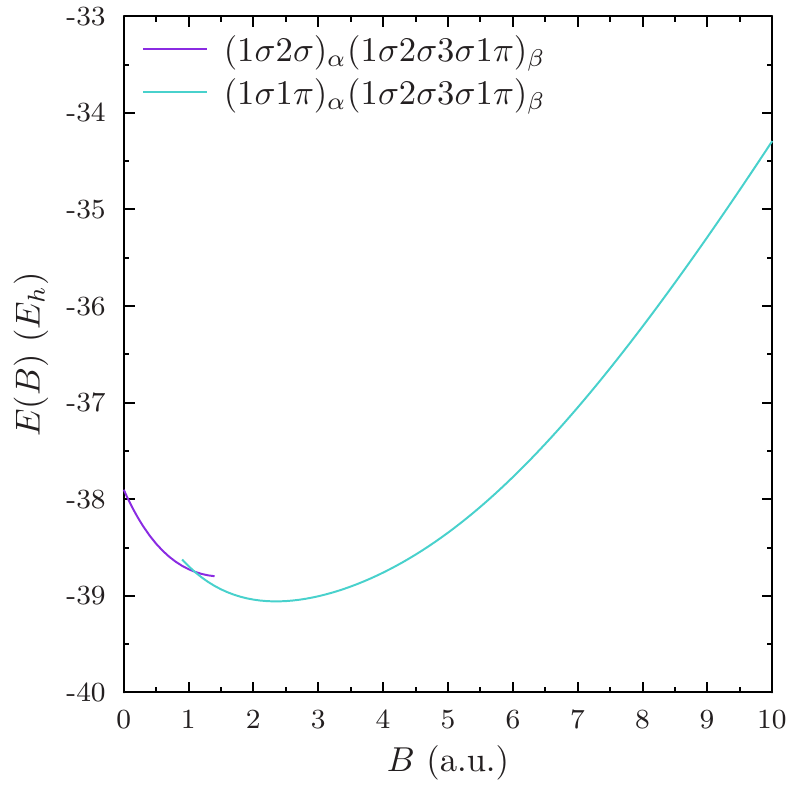}}
  \subfloat[][FEM]{\includegraphics[width=.49\textwidth]{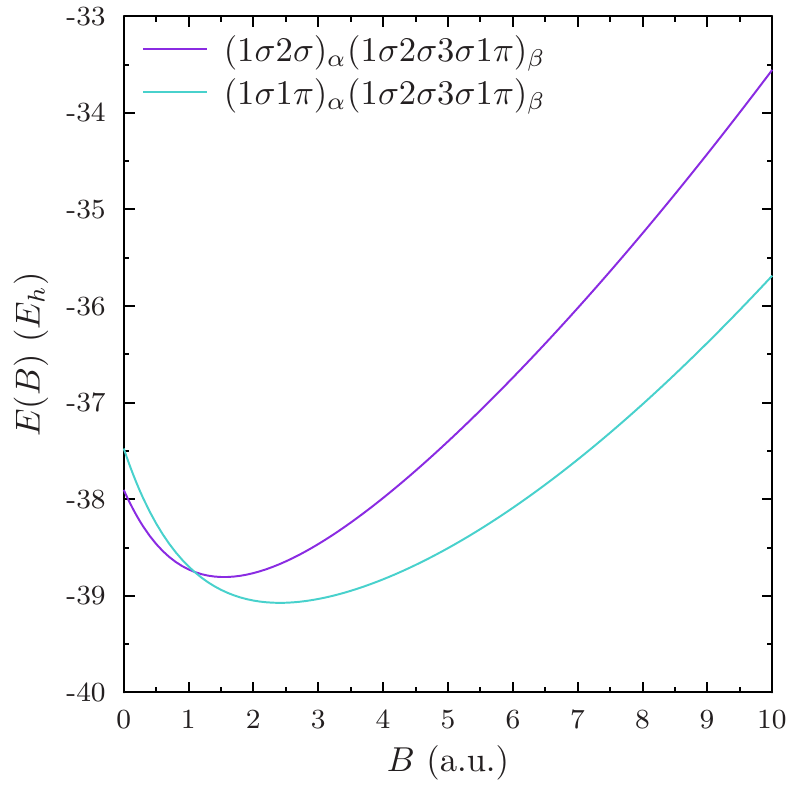}}
  \caption{\ce{CH+} triplet.\label{fig:CH+-3}}
\end{figure}

\begin{figure}
  \centering
  \subfloat[][GTO]{\includegraphics[width=.49\textwidth]{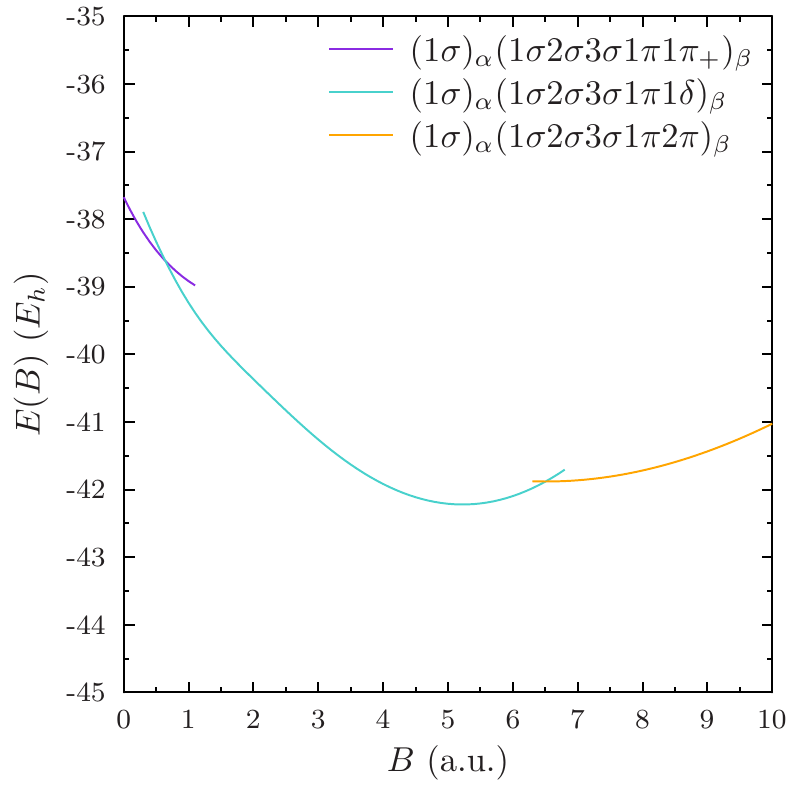}}
  \subfloat[][FEM]{\includegraphics[width=.49\textwidth]{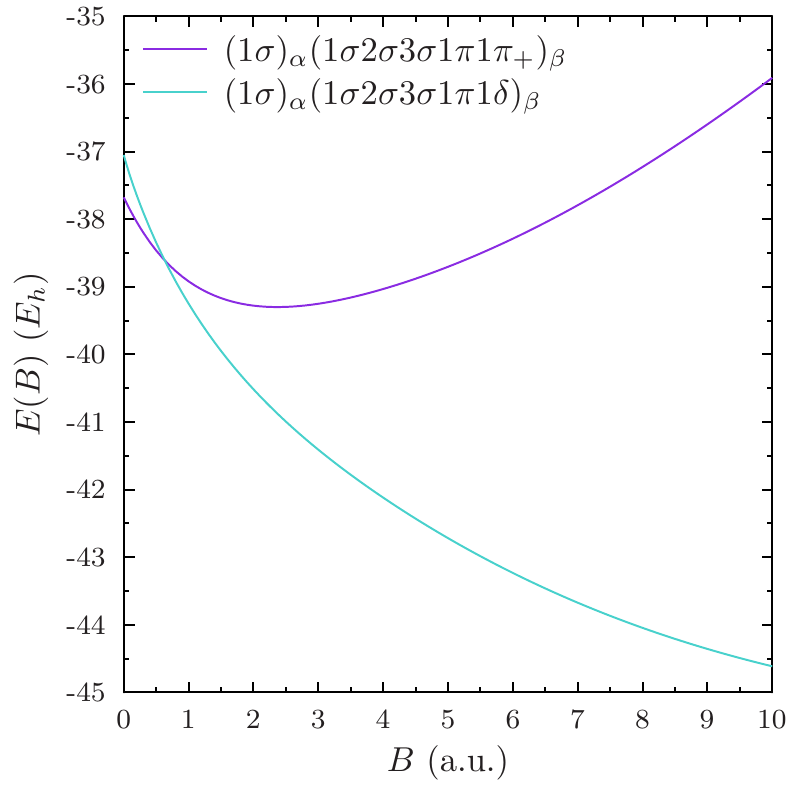}}
  \caption{\ce{CH+} quintet.\label{fig:CH+-5}}
\end{figure}

The total energy as a function of the magnetic field strength for all
states obtained in the calculations with the finite-element (FEM)
calculations with \textsc{HelFEM}, and the gauge-including
Gaussian-type orbital (GTO) calculations with \textsc{London} are
shown in \figrangeref{HH-1}{HH-3} for \ce{H2}, in
\figrangeref{HeH+-1}{HeH+-3} for \ce{HeH+}, in
\figrangeref{LiH-1}{LiH-5} for \ce{LiH}, in
\figrangeref{BeH+-1}{BeH+-5} for \ce{BeH+}, in
\figrangeref{BH-1}{BH-5} for \ce{BH}, and in
\figrangeref{CH+-1}{CH+-5} for \ce{CH+}. The resulting ground-state
configurations in the FEM and GTO calculations have been identified in
\tabref{gs-B-light} for \ce{H2}, \ce{HeH+}, and \ce{LiH}, and in
\tabref{gs-B-BeH+,gs-B-BH,gs-B-CH+} for \ce{BeH+}, \ce{BH} and
\ce{CH+}, respectively.

\clearpage

\clearpage

\begin{table}
\caption{Ground-state configurations for \ce{H2}, \ce{HeH+}, LiH, and
  \ce{BeH+} as a function of the magnetic field strength calculated at
  the Hartree--Fock level using gauge-including Gaussian-type orbitals
  (GTO) and finite element (FEM) basis sets.  The label X refers to
  symmetry-broken orbitals. Note that the orbitals are not sorted
  according to their energies.  \label{tab:gs-B-light}}

\begin{tabularx}{\linewidth}{llllc}
\hline
\hline
Molecule & spin state & basis & configuration & field strength ($B_0$) \\
\hline
\ce{H2} & singlet & FEM & $ 1\osa 1\osb $ & $ 0 \leq B \leq 10 $ \\
& & GTO & $ 1\osa 1\osb $ & $ 0 \leq B \leq 10  $ \\
& triplet & FEM & $ (1\si 2\si)_\be $ & $ 0 \leq B < 0.5 $ \\
&  &  & $ (1\si 1\pi)_\be $ & $ 0.5 \leq B < 5.2 $ \\
&  &  & $ (1\si 2\si)_\be $ & $ 5.2 \leq B \leq 10 $ \\
&  & GTO & $ (1\si 2\si)_\be $ & $ 0 \leq B < 0.9 $ \\
&  & & $ (1\si 1\pi)_\be $ & $ 0.9 \leq B < 2.2 $ \\
& & & $ (1\si 2\si)_\be $ & $ 2.2 \leq B \leq 10 $ \\
\\
\ce{HeH+} & singlet & FEM & $ 1\osa 1\osb $ & $ 0 \leq B \leq 10 $ \\
& & GTO & $ 1\osa 1\osb $ & $ 0 \leq B \leq 10 $ \\
& triplet & FEM &  $ (1\si 2\si)_\be $ & $ 0 \leq B \leq 10 $ \\
&  & GTO &  $ (1\si 2\si)_\be $ & $ 0 \leq B \leq 10 $ \\
\\
LiH & singlet & FEM & $ (1\si 2\si)_\al (1\si 2\si)_\be $ & $0 \leq B < 8.8  $ \\
& & & $ (1\si 1\pi)_\al (1\si 1\pi)_\be $ & $8.8 \leq B \leq 10  $ \\
& & GTO & $ (1\si 2\si)_\al (1\si 2\si)_\be $ & $0 \leq B < 9.6  $ \\
& & & $ (1\si X)_\al (1\si X)_\be $ & $ 9.6  \leq B \leq 10  $ \\
& triplet & FEM & $ 1\osa (1\si 2\si 3\si)_\be $ & $ 0 \leq B < 0.1  $ \\
& & & $ 1\osa (1\si 2\si 3\si)_\be $ & $ 0.1 \leq B \leq 10  $ \\
& & GTO &  $ 1\osa (1\si 2\si 3\si)_\be $ & $ 0 \leq B < 0.1  $ \\
& & & $ 1\osa (1\si 2\si 1\pi)_\be $ & $ 0.1  \leq B \leq 10  $ \\
& quintet & FEM & $ (1\si 2\si 3\si 1\pi)_\be $ & $ 0 \leq B < 0.1  $ \\
& & & $ (1\si 2\si 1\pi 3\si)_\be $ & $ 0.1 \leq B < 0.7  $ \\
& & & $ (1\si 2\si 1\pi 1\de)_\be $ & $ 0.7 \leq B < 2.0  $ \\
& & & $ (1\si 1\pi 2\si 1\de)_\be $ & $ 2.0 \leq B \leq 10  $ \\
& & GTO & $ (1\si 2\si 3\si 4\si)_\be $ & $ 0 \leq B < 0.01  $ \\
& & & $ (1\si 2\si 3\si 1\pi)_\be $ & $ 0.01  \leq B < 4.7  $ \\
& & & $ (1\si 2\si 1\pi 2\pi)_\be $ & $ 4.7  \leq B < 9.7  $ \\
& & & $ (1\si 2\si 3\si 1\pi)_\be $ & $ 9.7  \leq B \leq 10  $ \\
\hline
\hline
\end{tabularx}
\end{table}

\begin{table}
\caption{Ground-state configurations for \ce{BeH+} as a
  function of the magnetic field strength calculated at the
  Hartree--Fock level using gauge-including Gaussian-type orbitals
  (GTO) and finite element (FEM) basis sets. \label{tab:gs-B-BeH+}}

\begin{tabularx}{\linewidth}{lllc}
\hline
\hline
Spin state & basis & configuration & field strength ($B_0$) \\
\hline
singlet & FEM & $ (1\si 2\si)_\al (1\si 2\si)_\be $ & $ 0 \leq B < 1.4  $ \\
& & $ (1\si 1\pi)_\al (1\si 1\pi)_\be $ & $ 1.4 \leq B < 10  $ \\
& GTO & $ (1\si 2\si)_\al (1\si 2\si)_\be $ & $ 0 \leq B < 1.4  $ \\
& & $ (1\si 1\pi)_\al (1\si 1\pi)_\be $ & $ 1.4  \leq B \leq 10  $ \\
triplet & FEM & $ 1\osa (1\si 2\si 3\si)_\be $ & $ 0 \leq B < 0.2  $ \\
& & $ 1\osa (1\si 2\si 1\pi)_\be $ & $ 0.2 \leq B < 2.0  $ \\
& & $ 1\osa (1\si 1\pi 2\si)_\be $ & $ 2.0 \leq B < 10  $ \\
& GTO & $ 1\osa (1\si 2\si 3\si)_\be $ & $ 0 \leq B < 0.2  $ \\
& & $ 1\osa (1\si 2\si 1\pi)_\be $ & $ 0.2 \leq B < 2.1  $ \\
& & $ 1\osa (1\si 1\pi 2\si)_\be $ & $ 2.1  \leq B \leq 10  $ \\
quintet & FEM & $ (1\si 2\si 3\si 1\pi)_\be $ & $ 0 \leq B \leq 0.1  $ \\
& & $ (1\si 2\si 1\pi 3\si)_\be $ & $ 0.1 < B < 1.3  $ \\
& & $ (1\si 1\pi 2\si 3\si)_\be $ & $ 1.3 \leq B < 1.5  $ \\
& & $ (1\si 1\pi 2\si 1\de)_\be $ & $ 1.5 \leq B < 5.8  $ \\
& & $ (1\si 1\pi 1\de 2\si)_\be $ & $ 5.8 \leq B \leq 10  $ \\
& GTO & $ (1\si 2\si 3\si 1\pi)_\be $ & $ 0 \leq B < 0.2  $ \\
& & $ (1\si 2\si 1\pi 3\si)_\be $ & $ 0.2  \leq B < 1.3  $ \\
& & $ (1\si 1\pi 2\si 3\si)_\be $ & $ 1.3  \leq B \leq 10   $ \\
\hline
\hline
\end{tabularx}
\end{table}

\begin{table}
\caption{Ground-state configurations for \ce{BH} as a
  function of the magnetic field strength calculated at the
  Hartree--Fock level using gauge-including Gaussian-type orbitals
  (GTO) and finite element (FEM) basis sets.  ${\pi}_{+}$
  orbitals  have a higher energy in the presence
  of the magnetic field than at zero field. \label{tab:gs-B-BH}}

\begin{tabularx}{\linewidth}{lllc}
\hline
\hline
Spin state & basis & configuration & field strength ($B_0$) \\
\hline
singlet & FEM & $ (1\si 2\si 3\si)_\al (1\si 2\si 3\si)_\be $ & $ 0 \leq B < 0.3  $ \\
& & $ (1\si 2\si 1\pi)_\al (1\si 2\si 1\pi)_\be $ & $ 0.3 \leq B < 0.9  $ \\
& & $ (1\si 1\pi 2\si)_\al (1\si 1\pi 2\si)_\be $ & $ 0.9 \leq B \leq 10  $ \\
& GTO & $ (1\si 2\si 3\si)_\al (1\si 2\si 3\si)_\be $ & $ 0 \leq B < 0.3  $ \\
& & $ (1\si 2\si 1\pi)_\al (1\si 2\si 1\pi)_\be $ & $ 0.3  \leq B < 0.7  $ \\
& & $ (1\si 1\pi 2\si)_\al (1\si 1\pi 2\si)_\be $ & $ 0.7  \leq B \leq 10  $ \\
triplet & FEM & $ (1\si 2\si)_\al (1\si 2\si 3\si 1\pi)_\be $ & $ 0 \leq B < 0.3  $ \\
& & $ (1\si 2\si)_\al (1\si 2\si 1\pi 3\si)_\be $ & $ 0.3 \leq B < 0.9  $ \\
& & $ (1\si 2\si)_\al (1\si 1\pi 2\si 3\si)_\be $ & $ 0.9 \leq B < 1.2  $ \\
& & $ (1\si 1\pi)_\al (1\si 2\si 1\pi 3\si)_\be $ & $ 1.2 \leq B < 2.7  $ \\
& & $ (1\si 1\pi)_\al (1\si 1\pi 2\si 3\si)_\be $ & $ 2.7 \leq B < 3.6 $ \\
& & $ (1\si 1\pi)_\al (1\si 1\pi 2\si 2\pi)_\be $ & $ 3.6 \leq B \leq 10 $ \\
& GTO & $ (1\si 2\si)_\al (1\si 2\si 3\si 1\pi)_\be $ & $ 0 \leq B < 0.3  $ \\
& & $ (1\si 2\si)_\al (1\si 2\si 1\pi 3\si)_\be $ & $ 0.3  \leq B < 0.9  $ \\
& & $ (1\si 2\si)_\al (1\si 1\pi 2\si 3\si)_\be $ & $ 0.9  \leq B < 1.5  $ \\
& & $ (1\si 1\pi)_\al (1\si 2\si 1\pi 3\si)_\be  $ & $ 1.5  \leq B < 3.3  $ \\
& & $ (1\si 1\pi)_\al (1\si 1\pi 2\si 3\si)_\be  $ & $ 3.3  \leq B \leq 10  $ \\
quintet & FEM & $ 1\osa (1\si 2\si 3\si 1\pi 1\pi_+)_\be $ & $ 0 \leq B < 0.3  $ \\
& & $ 1\osa (1\si 2\si 1\pi 3\si 1\pi_+)_\be $ & $ B = 0.3  $ \\
& & $ 1\osa (1\si 2\si 1\pi 3\si 1\delta)_\be $ & $ 0.4 \leq B < 1.5   $ \\
& & $ 1\osa (1\si 1\pi 2\si 3\si 1\delta)_\be $ & $ 1.5 \leq B < 1.8   $ \\
& & $ 1\osa (1\si 1\pi 2\si 1\delta 3\si)_\be $ & $ 1.8 \leq B \leq 10   $ \\
& GTO & $ 1\osa (1\si 2\si 3\si 1\pi 1{\pi}_{+})_\be $ & $ 0 \leq B < 0.4  $ \\
& & $ 1\osa (1\si 2\si 3\si 4\si 1\pi)_\be $ & $ 0.4  \leq B < 0.7  $  \\
& & $ 1\osa (1\si 2\si 3\si 1\pi 2\pi)_\be $ & $ 0.7  \leq B < 1.5  $  \\
& & $ 1\osa (1\si 1\pi 2\si 3\si 1\de)_\be $ & $ 1.5  \leq B < 2.0  $  \\
& & $ 1\osa (1\si 1\pi 2\si 1\de 3\si)_\be $ & $ 2.0  \leq B < 3.3  $  \\
& & $ 1\osa (1\si 1\pi 2\si 3\si 1\de)_\be $ & $ 3.3  \leq B < 4.0  $  \\
& & $ 1\osa (1\si 2\si 3\si 4\si 1\pi)_\be $ & $ 4.0  \leq B \leq 10  $  \\
\hline
\hline
\end{tabularx}
\end{table}

\begin{table}
  \caption{Ground-state configurations for \ce{CH+} as a
    function of the magnetic field strength calculated at the
    Hartree--Fock level using gauge-including Gaussian-type orbitals
    (GTO) and finite element (FEM) basis sets.  ${\pi}_{+}$
    orbitals  have a higher energy in the presence
    of the magnetic field than at zero field. \label{tab:gs-B-CH+}}

  \begin{tabularx}{\linewidth}{lllc}
    \hline
    \hline
    Spin state & basis & configuration & field strength ($B_0$) \\
    \hline
    singlet & FEM & $ (1\si 2\si 3\si)_\al (1\si 2\si 3\si)_\be  $ & $ 0 \leq B < 0.3  $ \\
    & & $ (1\si 2\si 1\pi)_\al (1\si 2\si 1\pi)_\be  $ & $ 0.3 \leq B < 1.0 $ \\
    & & $ (1\si 1\pi 2\si)_\al (1\si 1\pi 2\si)_\be  $ & $ 1.0 \leq B \leq 10 $ \\
    & GTO & $ (1\si 2\si 3\si)_\al (1\si 2\si 3\si)_\be  $ & $ 0 \leq B < 0.3  $ \\
    & & $ (1\si 2\si 1\pi)_\al (1\si 2\si 1\pi)_\be $ & $ 0.3  \leq B < 1.0  $ \\
    & & $ (1\si 1\pi 2\si)_\al (1\si 1\pi 2\si)_\be $ & $ 1.0  \leq B \leq 10  $ \\
    triplet & FEM & $ (1\si 2\si)_\al (1\si 2\si 3\si 1\pi)_\be $ & $ 0 \leq B < 0.3  $ \\
    & & $ (1\si 2\si)_\al (1\si 2\si 1\pi 3\si)_\be $ & $ 0.3 \leq B < 1.0  $ \\
    & & $ (1\si 2\si)_\al (1\si 1\pi 2\si 3\si)_\be $ & $ B = 1.0  $ \\
    & & $ (1\si 1\pi)_\al (1\si 2\si 1\pi 3\si)_\be $ & $ 1.1 \leq B < 1.5  $ \\
    & & $ (1\si 1\pi)_\al (1\si 1\pi 2\si 3\si)_\be $ & $ 1.5 \leq B \leq 10  $ \\
    & GTO & $ (1\si 2\si)_\al (1\si 2\si 3\si 1\pi)_\be $ & $ 0 \leq B < 0.3  $ \\
    & & $ (1\si 2\si)_\al (1\si 2\si 1\pi 3\si)_\be $ & $ 0.3  \leq B < 1.0  $ \\
    & & $ (1\si 2\si)_\al (1\si 1\pi 2\si 3\si)_\be $ & $ 1.0  \leq B < 1.3  $ \\
    & & $ (1\si 1\pi)_\al (1\si 2\si 1\pi 3\si)_\be $ & $ 1.3  \leq B < 1.5  $ \\
    & & $ (1\si 1\pi)_\al (1\si 1\pi 2\si 3\si)_\be  $ & $ 1.5  \leq B \leq 10  $ \\
    quintet & FEM & $ 1\osa (1\si 2\si 3\si 1\pi 1\pi_+)_\be $ & $ 0 \leq B < 0.3  $ \\
    & & $ 1\osa (1\si 2\si 1\pi 3\si 1\pi_+)_\be $ & $ 0.3 \leq B < 0.7  $ \\
    & & $ 1\osa (1\si 2\si 1\pi 3\si 1\delta)_\be $ & $ 0.7 \leq B < 1.2  $ \\
    & & $ 1\osa (1\si 1\pi 2\si 3\si 1\delta)_\be $ & $ 1.2 \leq B < 3.0  $ \\
    & & $ 1\osa (1\si 1\pi 2\si 1\delta 3\si)_\be $ & $ 3.0 \leq B < 6.6  $ \\
    & & $ 1\osa (1\si 1\pi 1\delta 2\si 3\si)_\be $ & $ 6.6 \leq B \leq 10  $ \\
    & GTO & $ 1\osa (1\si 2\si 3\si 1\pi 1\pi_{+})_\be $ & $ 0 \leq B < 0.3  $ \\
    & & $ 1\osa (1\si 2\si 1\pi 3\si 1\pi_{+})_\be $ & $ 0.3 \leq B < 0.7  $ \\
    & & $ 1\osa (1\si 2\si 1\pi 3\si 1\de)_\be $ & $ 0.7 \leq B < 1.2  $ \\
    & & $ 1\osa (1\si 1\pi 2\si 3\si 1\de)_\be $ & $ 1.2 \leq B < 3.4  $ \\
    & & $ 1\osa (1\si 1\pi 2\si 1\de 3\si)_\be $ & $ 3.4 \leq B < 6.5  $ \\
    & & $ 1\osa (1\si 2\si 3\si 4\si 1\pi)_\be $ & $ 6.5  \leq B \leq 10  $ \\
    \hline
    \hline
  \end{tabularx}
\end{table}

\end{document}